\documentclass{svjour3}
\smartqed  % flush right qed marks, e.g. at end of proof
\usepackage{amssymb,amsmath}
\usepackage{graphicx}
\newcommand\rf[1]{(\ref{eq:#1})}
\newcommand\lab[1]{\label{eq:#1}}
\newcommand\nonu{\nonumber}
\newcommand\br{\begin{eqnarray}}
\newcommand\er{\end{eqnarray}}
\newcommand\be{\begin{equation}}
\newcommand\ee{\end{equation}}

\newcommand\foot[1]{\footnotemark\footnotetext{#1}}
\newcommand\lb{\lbrack}
\newcommand\rb{\rbrack}
\newcommand\llangle{\left\langle}
\newcommand\rrangle{\right\rangle}

\newcommand\llb{\left\lbrack}
\newcommand\rrb{\right\rbrack}

\newcommand\lcurl{\left\{}
\newcommand\rcurl{\right\}}
\renewcommand\({\left(}
\renewcommand\){\right)}
\newcommand\bv{\bigm\vert}               %%
\newcommand\bgv{\bigg\vert}              %%

\newcommand\bc{\begin{center}}
\newcommand\ec{\end{center}}

\newcommand\partder[2]{\frac{{\partial {#1}}}{{\partial {#2}}}}
\renewcommand\a{\alpha}
\renewcommand\b{\beta}

\renewcommand\d{\delta}

\newcommand\eps{\epsilon}
\newcommand\vareps{\varepsilon}
\newcommand\g{\gamma}
\newcommand\G{\Gamma}

\newcommand\h{\frac{1}{2}}
\renewcommand\k{\kappa}
\renewcommand\l{\lambda}
\renewcommand\L{\Lambda}
\newcommand\m{\mu}
\newcommand\n{\nu}

\newcommand\vp{\varphi}
\renewcommand\P{\Phi}
\newcommand\pa{\partial}

\newcommand\pr{\prime}

\renewcommand\r{\rho}
\newcommand\s{\sigma}
\renewcommand\S{\Sigma}
\renewcommand\t{\tau}
\renewcommand\th{\theta}

\newcommand\z{\zeta}

\newcommand\wti{\widetilde}
\newcommand\cA{{\mathcal A}}

\newcommand\cE{{\mathcal E}}
\newcommand\cF{{\mathcal F}}

\newcommand\cJ{{\mathcal J}}

\newcommand\cV{{\mathcal V}}

\newcommand{\ct}[1]{\cite{#1}}
\newcommand{\bib}[1]{\bibitem{#1}}
\newcommand\PRL[3]{\textsl{Phys. Rev. Lett.} \textbf{#1}, #3 (#2)}
\newcommand\NPB[3]{\textsl{Nucl. Phys.} \textbf{B#1}, #3 (#2)}

\newcommand\PRD[3]{\textsl{Phys. Rev.} \textbf{D#1}, #3 (#2)}

\newcommand\PLB[3]{\textsl{Phys. Lett.} \textbf{#1B}, #3 (#2)}
\newcommand\CQG[3]{\textsl{Class. Quantum Grav.} \textbf{#1}, #3 (#2)}

\newcommand\AoP[3]{\textsl{Ann. of Phys.} \textbf{#1}, #3 (#2)}

\newcommand\IJMPA[3]{\textsl{Int. J. Mod. Phys.} \textbf{A#1}, #3 (#2)}

\newcommand\Xdot{\stackrel{.}{X}}
\newcommand\xdot{\stackrel{.}{x}}

\newcommand\etadot{\stackrel{.}{\eta}}
\begin{document}

\title{Space-Time Compactification/Decompactification Transitions Via Lightlike Branes}

\titlerunning{Compactification/Decompactification Via Lightlike Branes}

% \author{Eduardo Guendelman \and Alexander Kaganovich \and Emil Nissimov \and 
% Svetlana Pacheva}
\author{Eduardo Guendelman \\ Alexander Kaganovich \\ 
Emil Nissimov \and Svetlana Pacheva}

\authorrunning{E. Guendelman, A. Kaganovich, E. Nissimov, S.Pacheva}

% \institute{E. Guendelman, A. Kaganovich
\institute{Eduardo Guendelman, Alexander Kaganovich
\at Department of Physics, Ben-Gurion University of the Negev, Beer-Sheva, Israel \\
\email{guendel@bgu.ac.il, alexk@bgu.ac.il}
\and
% E. Nissimov, S. Pacheva
Emil Nissimov, Svetlana Pacheva
\at Institute for Nuclear Research and Nuclear Energy,
Bulgarian Academy of Sciences, Sofia, Bulgaria  \\
\email{nissimov@inrne.bas.bg, svetlana@inrne.bas.bg}
}
\date{Received: date / Accepted: date}

\maketitle

\begin{abstract}
We consider Einstein-Maxwell-Kalb-Ramond gravity-matter system in bulk
space-time interacting self-consistently with two (widely separated) codimension-one
electrically charged {\em lightlike} branes. The lightlike brane dynamics is 
explicitly given by manifestly reparametrization invariant world-volume actions in two
equivalent dual to each other formulations (Polyakov-type and Nambu-Goto-type ones) 
proposed in our previous work. We find an explicit solution of the pertinent 
Einstein-Maxwell-Kalb-Ramond-lightlike-brane equations of motion describing a 
``two-throat'' wormhole-like space-time consisting of a ``left'' compactified 
Bertotti-Robinson universe connected to a ``middle'' non-compact 
Reissner-Nordstr{\"o}m-de-Sitter space-time region, which in turn is connected to
another ``right'' compactified Bertotti-Robinson universe. Each of the 
lightlike branes automatically occupies one of the ``throats'', so that they
dynamically induce a sequence of spontaneous space-time 
compactification/decompactification transitions.
%
% The aim of the present paper is two-fold. First we describe the Lagrangian
% dynamics of a recently proposed new class of {\em lightlike} $p$-branes and
% their interactions with bulk space-time gravity and electromagnetism in a
% self-consistent manner. Next, we discuss the role of {\em lightlike} branes
% as natural candidates for {\em wormhole} ``throats'' and exemplify the
% latter by presenting an explicit construction of a new type of asymmetric
% wormhole solution where the {\em lightlike} brane connects a ``right''
% universe with Reissner-Nordstr{\"o}m geometry to a ``left''
% Bertotti-Robinson universe with two compactified space dimensions.
%
%\keywords{traversable wormholes; non-Nambu-Goto lightlike branes;
% space-time compactification/decompactification}
\end{abstract}

%%%%%%%%%%%%%%%%%%%%%%%%%%%%%%%%%%%%%%%%%%%%%%%%%%%%%%%%%%%%%
%%%%%%%%%%%%%%%%%%%%%%%%%%%%%%%%%%%%%%%%%%%%%%%%%%%%%%%%%%%%%
\section{Introduction} \label{intro}

Lightlike branes (\textsl{LL-branes} for short) play an increasingly significant role 
in general relativity and modern non-perturbative string theory.
Mathematically they represent singular null hypersurfaces in Riemannian
space-time which provide dynamical description of various physically important 
cosmological and astrophysical phenomena such as:
(i) impulsive lightlike signals arising in cataclysmic astrophysical events 
(supernovae, neutron star collisions) \ct{barrabes-hogan}; 
(ii) dynamics of horizons in black hole physics -- the so called ``membrane paradigm''
\ct{membrane-paradigm};
(iii) the thin-wall approach to domain walls coupled to 
gravity \ct{Israel-66,Barrabes-Israel,Dray-Hooft}.

More recently, the relevance of \textsl{LL-branes} in the context of
non-perturbative string theory has also been recognized, specifically, as the so called
$H$-branes describing quantum horizons (black hole and cosmological)
\ct{kogan-01}, as Penrose limits of baryonic $D$-branes
\ct{mateos-02}, \textsl{etc} (see also Refs.\ct{nonperturb-string}).

A characteristic feature of the formalism for \textsl{LL-branes} in the
pioneering papers \ct{Israel-66,Barrabes-Israel,Dray-Hooft}
in the context of gravity and cosmology is that they have been exclusively
treated in a ``phenomenological'' manner, \textsl{i.e.}, without specifying
an underlying Lagrangian dynamics from which they may originate. As a
partial exception, in a more recent paper \ct{barrabes-israel-05} brane actions in 
terms of their pertinent extrinsic geometry
have been proposed which generically describe non-lightlike branes, whereas the 
lightlike branes are treated as a limiting case. 

On the other hand, in the last few years we have proposed in a series of papers 
\ct{LL-brane-main,inflation-all,our-WH,rot-WH} a new class of concise
manifestly reparametrization invariant world-volume Lagrangian actions, 
% among them -- {\em Weyl-conformally invariant} ones, 
providing a derivation from first principles of the \textsl{LL-brane} dynamics.
The following characteristic features of the new \textsl{LL-branes} drastically
distinguish them from ordinary Nambu-Goto branes: 

(a) They describe intrinsically lightlike modes, whereas Nambu-Goto branes describe
massive ones.

(b) The tension of the \textsl{LL-brane} arises as an {\em additional
dynamical degree of freedom}, whereas Nambu-Goto brane tension is a given
{\em ad hoc} constant. %%%%%%%%%%%
The latter characteristic feature significantly distinguishes our \textsl{LL-brane}
models from the previously proposed {\em tensionless} $p$-branes (for a review,
see Ref.\ct{lindstroem-etal}). The latter rather resemble $p$-dimensional continuous
distributions of independent massless point-particles without cohesion among the latter.

(c) Consistency of \textsl{LL-brane} dynamics in a spherically or axially
symmetric gravitational background of codimension one requires the presence
of an event horizon which is automatically occupied by the \textsl{LL-brane}
(``horizon straddling'' according to the terminology of Ref.\ct{Barrabes-Israel}).

(d) When the \textsl{LL-brane} moves as a {\em test} brane in spherically or 
axially symmetric gravitational backgrounds its dynamical tension exhibits 
exponential ``inflation/deflation'' time behavior \ct{inflation-all}
-- an effect similar to the ``mass inflation'' effect around black hole horizons
\ct{israel-poisson}. 

An intriguing novel application of \textsl{LL-branes} as natural self-consistent 
gravitational sources for {\em wormhole} space-times has been developed  in a series of 
recent papers \ct{our-WH,rot-WH,ER-bridge,varna-09}.

Before proceeding let us recall that the concept of ``wormhole space-time'' 
was born in the classic work
of Einstein and Rosen \ct{einstein-rosen}, where they considered matching along
the horizon of two identical copies of the exterior Schwarzschild space-time
region (subsequently called {\em Einstein-Rosen ``bridge''}).
Another corner stone in wormhole physics is the seminal work of Morris and
Thorne \ct{morris-thorne}, who studied for the first time {\em traversable Lorentzian
wormholes}.
%%%%%%%%%%%%%%%%%%%%%%%%%%%%%%%%%%%
% \vspace{.2in}

% %%%%%%%%%%%%%%%%%%%%%%%%%%%%%%%%%%%%%%%%%%%%%%%%%%%%%%%%%%%%%%%%%%
% \%\%\%\%\%\%\%\%\%\%\%\%\%\%\%\%\%\%\%\%\%\%\%\%\%\%\%\%\%\%\%\%\%

% \textbf{Morris-Thorne \& generalizations, Visser-Hochberg} 

% \textbf{violation of null-energy condition}

% \textbf{More recent developments - Lemos, Lobo, Sushkov,
% Zaslavski. Maeda (null hypersurface matching)}

% \textbf{Class of ``thin-shell'' wormholes}

% \textbf{Holography and wormholes in 2+1 dimensions, Skenderis etal}

% \%\%\%\%\%\%\%\%\%\%\%\%\%\%\%\%\%\%\%\%\%\%\%\%\%\%\%\%\%\%\%\%\%
% %%%%%%%%%%%%%%%%%%%%%%%%%%%%%%%%%%%%%%%%%%%%%%%%%%%%%%%%%%%%%%%%%%

% \vspace{.2in}
%%%%%%%%%%%%%%%%%%%%%%%%%%%%%%%%%%%
In what follows, when discussing wormholes we will have in mind the physically 
important class of ``thin-shell'' traversable Lorentzian wormholes first introduced by
Visser \ct{visser-thin,visser-book}.
For a comprehensive review of wormhole space-times, see Refs.\ct{visser-book,WH-rev}.

In our earlier work \ct{our-WH,rot-WH,ER-bridge,varna-09} we have constructed
various types of wormhole solutions in self-consistent systems of bulk gravity 
and bulk gauge fields (Maxwell and Kalb-Ramond) coupled to \textsl{LL-branes} 
where the latter provide the appropriate stress energy tensors, electric
currents and dynamically generated space-varying cosmological constant terms
consistently derived from well-defined world-volume \textsl{LL-brane} Lagrangian actions.

The original Einstein-Rosen ``bridge'' manifold \ct{einstein-rosen} appears as a 
particular case of the construction of spherically symmetric wormholes produced by
\textsl{LL-branes} as gravitational sources occupying the wormhole throats
(Refs.\ct{ER-bridge,rot-WH}). 
% Therefore, the introduction of \textsl{LL-brane} coupling to gravity 
% as sources of wormhole space-times 
% brings the original Einstein-Rosen construction in Ref.\ct{einstein-rosen} 
% to a consistent completion
Thus, we are lead to the important conclusion
that consistency of Einstein equations of motion yielding the original Einstein-Rosen 
``bridge'' as well-defined solution necessarily requires the presence of 
\textsl{LL-brane} energy-momentum tensor as a source on the right hand 
side\foot{The crucial role of the presence of the 
\textsl{LL-brane} gravitational source producing the Einstein-Rosen ``bridge'' 
was not recognized in the original classic paper \ct{einstein-rosen}.
On the other hand, the failure of the original Einstein-Rosen ``bridge'' metric 
to satisfy the vacuum Einstein equations at the matching hypersurface -- the
Schwarzschild horizon, which now serves as wormhole ``throat'', 
has been noticed in ref.\ct{einstein-rosen}, where in Eq.(3a)
the authors multiply Ricci tensor by an appropriate power of the determinant
of the ``bridge'' metric vanishing at the ``throat'' so as to enforce
fulfillment of the vacuum Einstein equations everywhere, including at the
``throat''. To avoid confusion, let us particularly emphasize that 
here we consider the Einstein-Rosen ``bridge'' in  its original formulation in 
Ref.\ct{einstein-rosen} as a four-dimensional space-time manifold consisting of two 
copies of the exterior Schwarzschild space-time region matched along the horizon.
On the other hand, the nomenclature of ``Einstein-Rosen bridge'' in several standard 
textbooks (\textsl{e.g.} Ref.\ct{MTW}) 
uses the Kruskal-Szekeres manifold. The latter notion of ``Einstein-Rosen bridge''
is not equivalent to the original construction in Ref.\ct{einstein-rosen}. Namely, 
the two regions in Kruskal-Szekeres space-time corresponding to the outer 
Schwarzschild space-time region ($r>2m$) and labeled $(I)$ and $(III)$ 
in Refs.\ct{MTW} are generally
{\em disconnected} and share only a two-sphere (the angular part) as a common border
($U=0, V=0$ in Kruskal-Szekeres coordinates), whereas in the original Einstein-Rosen
``bridge'' construction the boundary between the two identical copies of the
outer Schwarzschild space-time region ($r>2m$) is a three-dimensional hypersurface
($r=2m$).}.
%%%%%%%%%%%%%%%%%%%%%%%%%%%%%%%%%%%

More complicated examples of spherically and axially symmetric wormholes with
Reissner-Nordstr{\"o}m and rotating cylindrical geometry, respectively,
have been explicitly constructed via \textsl{LL-branes} in Refs.\ct{our-WH,rot-WH}. 
Namely, two copies of the exterior space-time region (\textsl{i.e.}, the space-time 
region beyond the respective outer event horizon) of a Reissner-Nordstr{\"o}m or
rotating cylindrical black hole, respectively, 
are matched via \textsl{LL-brane} along what used to be the outer horizon of the
respective full black hole space-time manifold. In this way one
obtains a wormhole solution which combines the features of the Einstein-Rosen
``bridge'' on the one hand (with wormhole throat at horizon), and the features of
Misner-Wheeler wormholes \ct{misner-wheeler}, \textsl{i.e.}, exhibiting the so called 
``charge without charge'' phenomenon\foot{Misner and Wheeler \ct{misner-wheeler}
realized that wormholes connecting two asymptotically flat space times provide 
the possibility of ``charge without charge'', \textsl{i.e.}, electromagnetically
non-trivial solutions where the lines of force of the electric field flow from one 
universe to the other without a source and giving the impression of being 
positively charged in one universe and negatively charged in the other universe.},
on the other hand.

The results of Refs.\ct{our-WH,rot-WH} have been further extended in \ct{varna-09,BR-WH}
to the case of {\em asymmetric} wormholes, describing two ``universes'' with different 
(spherically symmetric) geometries of black hole type connected via a ``throat'' 
materialized by the pertinent gravitational source -- an electrically charged 
\textsl{LL-brane}, sitting on their common horizon. In \ct{varna-09} it has been shown 
that as a result of the well-defined world-volume dynamics of the \textsl{LL-brane}
coupled self-consistently to gravity and bulk space-time gauge fields, it creates a
``left universe'' comprising the exterior Schwarzschild-de-Sitter space-time
region beyond the Schwarzschild horizon and where the cosmological constant is 
dynamically generated, and a ``right universe'' comprising the exterior 
Reissner-Nordstr{\"o}m region beyond the outer Reissner-Nordstr{\"o}m horizon
with dynamically generated Coulomb field-strength. Both ``universes'' are
glued together by the \textsl{LL-brane} occupying their common horizon.
Similarly, the \textsl{LL-brane} can dynamically generate a non-zero
cosmological constant in the ``right universe'', in which case it connects a
purely Schwarzschild ``left universe'' with a Reissner-Nordstr{\"o}m-de-Sitter
``right universe''. Furthermore, in ref.\ct{BR-WH} another new type of wormhole
solution to Einstein-Maxwell equations has been constructed, which describes a 
``right universe'' comprising the exterior Reissner-Nordstr{\"o}m space-time region 
beyond the outer Reissner-Nordstr{\"o}m horizon, connected through a
``throat'' materialized by a \textsl{LL-brane} with a ``left universe''
being a Bertotti-Robinson space-time with two compactified spatial dimensions
\ct{BR}\foot{Originally space-time solution with two compactified spatial dimensions
was found by Levi-Civita \ct{LC}; see also \ct{lapedes-78}.}. Thus, the 
\textsl{LL-brane} has been shown to dynamically induce space-time compactification.

Let us note that previously the junction of a compactified space-time (of
Bertotti-Robinson type) to an uncompactified space-time through a wormhole
has been studied in a different setting using {\em timelike} matter on the 
junction hypersurface \ct{eduardo-GRG}. Recently \ct{halilsoy} a general class of
solutions of Einstein-Maxwell-Dilaton system have been found describing an
interpolation between Reissner-Nordstr{\"o}m and Bertotti-Robinson space-times.
Also, in a different context a 
string-like (flux tube) object with similar features to Bertotti-Robinson
solution has been constructed \ct{dzhunu} which interpolates between
uncompactified space-time regions.
%%%%%%%%%%%%%%%%%%%%%%%%%%%%%%%%%%%%%%%

In the present paper we will further broaden the application of \textsl{LL-branes}
in the context of wormhole physics by constructing a {\em ``two-throat''} wormhole-type
solution to Einstein-Maxwell system self-consistently interacting with {\em two
widely separated} electrically charged \textsl{LL-branes} describing a
sequence of a decompactification and a subsequent compactification
transitions interpolating between two Bertotti-Robinson universes with
different sizes of the compactified dimensions,
\textsl{i.e.}, a sort of space-time compactification ``kink'' solution.

% %%%%%%%%%%%%%%%%%%%%%%%%%%%%%%%%%%%%%%%%%%%%%%%%%%%%%%%%%%%%%%%%%%
% \%\%\%\%\%\%\%\%\%\%\%\%\%\%\%\%\%\%\%\%\%\%\%\%\%\%\%\%\%\%\%\%\%

% \textbf{Relation with Eduardo's WIS-91-27 ``Wormholes and Compactification''}

% \textbf{Dzhunushaliev's ``flux-tube'' space-times}
% % V. Dzhunushaliev, Class.Quant.Grav.20:2407,2003 (gr-qc/0301118),
% % V. Dzhunushaliev, Gen.Rel.Grav.35:1481,2003 ( gr-qc/0301046),
% % V. Dzhunushaliev, Class.Quant.Grav.19:4817,2002 (gr-qc/0205055), 
% % V. Dzhunushaliev , U. Kasper, D. Singleton, Phys.Lett.B479:249,2000 (gr-qc/9910092).

% \%\%\%\%\%\%\%\%\%\%\%\%\%\%\%\%\%\%\%\%\%\%\%\%\%\%\%\%\%\%\%\%\%
% %%%%%%%%%%%%%%%%%%%%%%%%%%%%%%%%%%%%%%%%%%%%%%%%%%%%%%%%%%%%%%%%%%

% \vspace{.2in}

The presentation of the material in the paper is as follows.
Sections \ref{sec:2} and \ref{sec:3} contain the basics of our formalism.
In Section \ref{sec:2} we review the
reparametrization-invariant world-volume Lagrangian formulation of \textsl{LL-branes}
in both the Polyakov-type and Nambu-Goto-type forms. In Section \ref{sec:3} we briefly
describe the main properties of \textsl{LL-brane} dynamics in spherically symmetric
gravitational backgrounds stressing particularly on the ``horizon straddling'' 
phenomenon and the dynamical cosmological constant generation. Section \ref{sec:4}
presents the systematic Lagrangian formulation of bulk Einstein-Maxwell-Kalb-Ramond
system self-consistently interacting with two distantly separated charged 
codimension-one \textsl{LL-branes}. Section \ref{sec:5}
contains our principal result -- the explicit construction of a new asymmetric
{\em ``two-throat''} wormhole solution of the above coupled 
gravity/gauge-\textsl{LL-brane} system describing: 
% self-consistent Einstein-Maxwell system
% interacting with two electrically charged \textsl{LL-branes} describing:

(i) Bertotti-Robinson compactified ``left universe'' connected via the first 
\textsl{LL-brane} to the

(ii) uncompactified ``middle universe'' comprising the intermediate space-time region of 
Reissner-Nordstr{\"o}m-de-Sitter universe between the middle (outer Reissner-Nordstr{\"o}m)
horizon and the outmost (de Sitter) horizon, connected in turn via the second
\textsl{LL-brane} to 

(iii) another Bertotti-Robinson ``right universe'' with {\em different} (w.r.t. (i)) 
size of the compactified dimensions.

In the concluding Section \ref{sec:6} we briefly discuss the issue of traversability of
the above wormhole solution as well as possible generalizations to higher space-time 
dimensions.

%%%%%%%%%%%%%%%%%%%%%%%%%%%%%%%%%%%%%%%%%%%%%%%%%%%%%%%%%%%%%%%%%%
%%%%%%%%%%%%%%%%%%%%%%%%%%%%%%%%%%%%%%%%%%%%%%%%%%%%%%%%%%%%%%%%%%
\section{World-Volume Formulation of Lightlike Brane Dynamics}
\label{sec:2}
%%%%%%%%%%%%%%%%%%%%%%%%%%%%%%%%%%%%%%%%%%%%%%%%%%%%%%%%%%%%%%%%%%
\subsection{Polyakov-Type Formulation}
\label{sec:2a}
There exist two equivalent {\em dual to each other} manifestly
reparametrization invariant world-volume Lagrangian formulations of 
\textsl{LL-branes} \ct{LL-brane-main,inflation-all,our-WH,rot-WH,ER-bridge,reg-BH}. 
First, let us consider the Polyakov-type formulation where the \textsl{LL-brane} 
world-volume action is given as:
\be
S_{\mathrm{Pol}} = \int d^{p+1}\s\,\P\llb -\h\g^{ab} g_{ab} + L\!\( F^2\)\rrb \; .
\lab{LL-action}
\ee
Here the following notions and notations are used:

\begin{itemize}
\item
(a) $\P$ is an alternative non-Riemannian integration measure density (volume form) 
on the $p$-brane world-volume manifold:
% \br
% \P \equiv \frac{1}{(p+1)!} 
% \vareps^{a_1\ldots a_{p+1}} H_{a_1\ldots a_{p+1}}(B) \;\; ,
% \lab{mod-measure-p} \\
% H_{a_1\ldots a_{p+1}}(B) = (p+1) \pa_{[a_1} B_{a_2\ldots a_{p+1}]} \; ,
% \lab{H-def}
% \er
\be
\P \equiv \frac{1}{(p+1)!} 
\vareps^{a_1\ldots a_{p+1}} H_{a_1\ldots a_{p+1}}(B) \quad ,\quad
H_{a_1\ldots a_{p+1}}(B) = (p+1) \pa_{[a_1} B_{a_2\ldots a_{p+1}]} \; ,
\lab{mod-measure-p}
\ee
instead of the usual $\sqrt{-\g}$. Here $\vareps^{a_1\ldots a_{p+1}}$ is the
alternating symbol ($\vareps^{0 1\ldots p} = 1$), $\g_{ab}$ ($a,b=0,1,{\ldots},p$)
indicates the intrinsic Riemannian metric on the world-volume, and
$\g = \det\Vert\g_{ab}\Vert$.
$H_{a_1\ldots a_{p+1}}(B)$ denotes the field-strength of an auxiliary
world-volume antisymmetric tensor gauge field $B_{a_1\ldots a_{p}}$ of rank $p$.
As a special case one can build $H_{a_1\ldots a_{p+1}}$ in terms of 
$p+1$ auxiliary world-volume scalar fields $\lcurl \vp^I \rcurl_{I=1}^{p+1}$:
% $\P \equiv \frac{1}{(p+1)!} \vareps_{I_1\ldots I_{p+1}}
% \vareps^{a_1\ldots a_{p+1}} \pa_{a_1} \vp^{I_1}\ldots \pa_{a_{p+1}} \vp^{I_{p+1}}$.
\be
% \P \equiv \frac{1}{(p+1)!} \vareps_{I_1\ldots I_{p+1}}
% \vareps^{a_1\ldots a_{p+1}} \pa_{a_1} \vp^{I_1}\ldots \pa_{a_{p+1}} \vp^{I_{p+1}} \;.
H_{a_1\ldots a_{p+1}} = \vareps_{I_1\ldots I_{p+1}}
\pa_{a_1} \vp^{I_1}\ldots \pa_{a_{p+1}} \vp^{I_{p+1}} \;.
\lab{mod-measure-p-scalar}
\ee
Note that $\g_{ab}$ is {\em independent} of the auxiliary world-volume fields
$B_{a_1\ldots a_{p}}$ or $\vp^I$.
% Note that $\g_{ab}$ is {\em independent} of $B_{a_1\ldots a_{p}}$.
The alternative non-Riemannian volume form \rf{mod-measure-p}
has been first introduced in the context of modified standard (non-lightlike) string and
$p$-brane models in Refs.\ct{mod-measure}.

\item
(b) $X^\m (\s)$ are the $p$-brane embedding coordinates in the bulk
$D$-dimensional space time with bulk Riemannian metric
$G_{\m\n}(X)$ with $\m,\n = 0,1,\ldots ,D-1$; 
$(\s)\equiv \(\s^0 \equiv \t,\s^i\)$ with $i=1,\ldots ,p$;
$\pa_a \equiv \partder{}{\s^a}$.

\item
(c) $g_{ab}$ is the induced metric on world-volume:
\be
g_{ab} \equiv \pa_a X^{\m} \pa_b X^{\n} G_{\m\n}(X) \; ,
\lab{ind-metric}
\ee
which becomes {\em singular} on-shell (manifestation of the lightlike nature, 
cf. second Eq.\rf{on-shell-singular} below).

\item
(d) $L\!\( F^2\)$ is the Lagrangian density of another
auxiliary $(p-1)$-rank antisymmetric tensor gauge field $A_{a_1\ldots a_{p-1}}$
on the world-volume with $p$-rank field-strength and its dual:
\be
F_{a_1 \ldots a_{p}} = p \pa_{[a_1} A_{a_2\ldots a_{p}]} \quad ,\quad
F^{\ast a} = \frac{1}{p!} \frac{\vareps^{a a_1\ldots a_p}}{\sqrt{-\g}}
F_{a_1 \ldots a_{p}}  \; .
\lab{p-rank}
\ee
$L\!( F^2)$ is {\em arbitrary} function of $F^2$ with the short-hand notation:
% $F^2 \equiv F_{a_1 \ldots a_{p}} F_{b_1 \ldots b_{p}} \g^{a_1 b_1} \ldots \g^{a_p b_p}$.
\be
F^2 \equiv F_{a_1 \ldots a_{p}} F_{b_1 \ldots b_{p}} 
\g^{a_1 b_1} \ldots \g^{a_p b_p} \; .
\lab{F2-id}
\ee
\end{itemize}

% Let us note the simple identity:
% \be
% F_{a_1 \ldots a_{p-1}b}F^{\ast b} = 0 \; ,
% \lab{F-id}
% \ee
% which will play a crucial role in the sequel.

Rewriting the action \rf{LL-action} in the following equivalent form:
\be
S_{\mathrm{Pol}} = - \int d^{p+1}\!\s \,\chi \sqrt{-\g}
\Bigl\lb \h \g^{ab} g_{ab} - L\!\( F^2\)\Bigr\rb
\;\; , \; \chi \equiv \frac{\P}{\sqrt{-\g}}
\lab{LL-action-chi}
\ee
with $\P$ the same as in \rf{mod-measure-p},
we find that the composite field $\chi$ plays the role of a {\em dynamical
(variable) brane tension}\foot{The notion of dynamical brane tension has previously 
appeared in different contexts in Refs.\ct{townsend-etal}.}.

Let us now consider the equations of motion corresponding to \rf{LL-action} 
w.r.t. $B_{a_1\ldots a_{p}}$:
\be
\pa_a \Bigl\lb \h \g^{cd} g_{cd} - L(F^2)\Bigr\rb = 0 \quad \longrightarrow \quad
\h \g^{cd} g_{cd} - L(F^2) = M  \; ,
\lab{phi-eqs}
\ee
where $M$ is an arbitrary integration constant. The equations of motion w.r.t.
$\g^{ab}$ read:
\be
g_{ab} - 2 F^2 L^{\pr}(F^2) \llb\g_{ab} 
- \frac{F^{*}_a F^{*}_b}{F^{*\, 2}}\rrb = 0  \; ,
\lab{gamma-eqs}
\ee
where $F^{*\, a}$ is the dual field strength \rf{p-rank}. Eqs.\rf{gamma-eqs} can be 
viewed as $p$-brane analogues of the string Virasoro constraints.

\vspace{.2in}
\textbf{Remark 1.}
Before proceeding, let us mention that both the auxiliary world-volume field 
$B_{a_1\ldots a_{p}}$ entering the non-Riemannian integration measure density
\rf{mod-measure-p}, as well as the intrinsic world-volume metric $\g_{ab}$ are
{\em non-dynamical} degrees of freedom in the action \rf{LL-action},
or equivalently, in \rf{LL-action-chi}. Indeed, there are no (time-)derivatives
w.r.t. $\g_{ab}$, whereas the action \rf{LL-action} (or \rf{LL-action-chi}) is
{\em linear} w.r.t. the velocities $\pa_0 B_{a_1\ldots a_{p}}$. Thus,
\rf{LL-action} is a constrained dynamical system, \textsl{i.e.}, a system with 
gauge symmetries including the gauge symmetry under world-volume
reparametrizations, and both Eqs.\rf{phi-eqs}--\rf{gamma-eqs} are in fact
{\em non-dynamical constraint} equations (no second-order time derivatives
present). Their meaning as constraint
equations is best understood within the framework of the Hamiltonian formalism 
for the action \rf{LL-action}. The latter can be developed
in strict analogy with the Hamiltonian formalism for a simpler class of
modified {\em non-lightlike} $p$-brane models based on the alternative non-Riemannian
integration measure density \rf{mod-measure-p}, which was previously proposed 
in Ref.\ct{m-string} (for details, we refer to Sections 2 and 3 of 
Ref.\ct{m-string}). In particular, Eqs.\rf{gamma-eqs} can be viewed as $p$-brane 
analogues of the string Virasoro constraints.

\vspace{.2in}
Taking the trace in \rf{gamma-eqs} and comparing with \rf{phi-eqs} 
implies the following crucial relation for the Lagrangian function $L\( F^2\)$: 
\be
% L\!\( F^2\) - p F^2 L^\pr\!\( F^2\) + M = 0 \; ,
L\!\( F^2\) - p F^2 L^\pr\!\( F^2\) + M = 0 \quad \to \quad 
F^2 = F^2 (M) = \mathrm{const}\; ,
\lab{L-eq}
\ee
% $L\!\( F^2\) - p F^2 L^\pr\!\( F^2\) + M = 0$, 
which determines $F^2$ \rf{F2-id} on-shell as certain function of the integration
constant $M$ \rf{phi-eqs}: %, \textsl{i.e.} $F^2 = F^2 (M) = \mathrm{const}$.
\be
F^2 = F^2 (M) = \mathrm{const} \; .
\lab{F2-const}
\ee
% In Eq.\rf{L-eq} 
Here and below $L^\pr(F^2)$ denotes derivative of $L(F^2)$ w.r.t. the 
argument $F^2$.

The next and most profound consequence of Eqs.\rf{gamma-eqs} is that the induced 
metric \rf{ind-metric} on the world-volume of the $p$-brane model \rf{LL-action} 
is {\em singular} on-shell (as opposed to the induced metric in the case of 
ordinary Nambu-Goto branes):
\be
% g_{ab}F^{*\, b}=0 \; .
g_{ab}F^{*\, b} \equiv \pa_a X^\m G_{\m\n} \(\pa_b X^\n F^{*\, b}\) =0 \; .
\lab{on-shell-singular}
\ee
Eq.\rf{on-shell-singular} is the manifestation of the {\em lightlike} nature
of the $p$-brane model \rf{LL-action} (or \rf{LL-action-chi}),
namely, the tangent vector to the world-volume $F^{*\, a}\pa_a X^\m$
is {\em lightlike} w.r.t. metric of the embedding space-time.

Further, the equations of motion w.r.t. world-volume gauge field 
$A_{a_1\ldots a_{p-1}}$ (with $\chi$ as defined in \rf{LL-action-chi}
and accounting for the constraint \rf{F2-const}) read:
\be
\pa_{[a}\( F^{\ast}_{b]}\, \chi\) = 0  \; .
\lab{A-eqs-0}
\ee

Finally, the $X^\m$ equations of motion produced by the \rf{LL-action} read:
\be
\pa_a \(\chi \sqrt{-\g} \g^{ab} \pa_b X^\m\) + 
\chi \sqrt{-\g} \g^{ab} \pa_a X^\n \pa_b X^\l \G^\m_{\n\l} = 0  \;
\lab{X-eqs-0}
\ee
where $\G^\m_{\n\l}=\h G^{\m\k}\(\pa_\n G_{\k\l}+\pa_\l G_{\k\n}-\pa_\k G_{\n\l}\)$
is the Christoffel connection for the external metric.

%%%%%%%%%%%%%%%%%%%%%%%%%%%%%%%%%%%%%%%%%%%%%%%%%%%%%%%%%%%%%%%%%%
\subsection{Nambu-Goto-Type Formulation}
\label{sec:2b}

Eq.\rf{A-eqs-0} allows us to introduce the dual ``gauge'' potential $u$ (dual
w.r.t. world-volume gauge field $A_{a_1\ldots a_{p-1}}$ \rf{p-rank}) :
\be
F^{\ast}_{a} = c_p\, \frac{1}{\chi} \pa_a u \quad ,\quad
c_p = \mathrm{const} \; . % \; ,
\lab{u-def}
\ee
% where the constant $c_p$ can be chosen in the following form which will
% simplify the matching with the numerical coefficients in the dual
% Nambu-Goto-type formulation (see Eq.\rf{LL-action-NG-A} below):
% \be
% c_p^2 = \frac{1}{p!} F^2 (2 a_0)^{p-2}\bv_{F^2=F^2(M)} \quad \mathrm{with}
% \;\;\; a_0 \equiv F^2 L^{\pr}\( F^2\)\bv_{F^2=F^2(M)} = \mathrm{const} \; .
% \lab{a0-eq}
% \ee
Relation \rf{u-def} enables us to rewrite Eq.\rf{gamma-eqs} (the lightlike constraint)
in terms of the dual potential $u$ in the form:
\be
\g_{ab} = \frac{1}{b_0}\, g_{ab} - \frac{b_0^{p-2}}{\chi^2}\,\pa_a u \pa_b u
\quad \mathrm{with}
\;\;\; b_0 \equiv 2 F^2 L^{\pr}\( F^2\)\bv_{F^2=F^2(M)} = \mathrm{const} \; .
\lab{gamma-eqs-u}
\ee
% ($L^\pr(F^2)$ denotes derivative of $L(F^2)$ w.r.t. the argument $F^2$).
From \rf{u-def} % and \rf{F2-const} 
we obtain the relation: 
\be
\chi^2 = -b_0^{p-2} \g^{ab} \pa_a u \pa_b u \; ,
\lab{chi2-eq}
\ee
and the Bianchi identity $\nabla_a F^{\ast\, a}=0$ becomes:
\be
\pa_a \Bigl( \frac{1}{\chi}\sqrt{-\g} \g^{ab}\pa_b u\Bigr) = 0  \; .
\lab{Bianchi-id}
\ee

It is straightforward to prove that the system of equations \rf{X-eqs-0},
\rf{Bianchi-id} and \rf{chi2-eq} for $\( X^\m,u,\chi\)$, which are equivalent to the 
equations of motion \rf{phi-eqs}--\rf{A-eqs-0},\rf{X-eqs-0} resulting from the 
original Polyakov-type \textsl{LL-brane} action \rf{LL-action}, can be equivalently 
derived from the following {\em dual} Nambu-Goto-type world-volume action:
\be
S_{\rm NG} = - \int d^{p+1}\s \, T 
\sqrt{\bgv\, \det\Vert g_{ab} - \eps \frac{1}{T^2}\pa_a u \pa_b u\Vert\,\bgv}
\quad ,\quad \eps = \pm 1 \; .
\lab{LL-action-NG-A}
\ee
Here again $g_{ab}$ indicates the induced metric on the world-volume \rf{ind-metric}
and $T$ is dynamical variable tension simply proportional to $\chi$ 
(see \rf{Pol-NG-rel} below).
The choice of the sign in \rf{LL-action-NG-A} does not have physical effect
because of the non-dynamical nature of the $u$-field.

The corresponding equations of motion w.r.t. $X^\m$, $u$ and $T$ read accordingly:
\br
\pa_a \( T \sqrt{|{\wti g}|} {\wti g}^{ab}\pa_b X^\m\)
+ T \sqrt{|{\wti g}|} {\wti g}^{ab} \pa_a X^\l \pa_b X^\n \G^\m_{\l\n} = 0 \; ,
\lab{X-eqs-NG}\\
% \pa_a \(\frac{1}{T} \sqrt{|{\wti g}|} {\wti g}^{ab}\pa_b u\) = 0 \; ,
% \lab{u-eqs-NG} \\
% T^2 + \eps {\wti g}^{ab}\pa_a u \pa_b u = 0 \; ,
% \lab{T-eq-NG}
\pa_a \(\frac{1}{T} \sqrt{|{\wti g}|} {\wti g}^{ab}\pa_b u\) = 0 \quad ,\quad
T^2 + \eps {\wti g}^{ab}\pa_a u \pa_b u = 0 \; ,
\lab{u-T-eqs-NG}
\er
where we have introduced the convenient notations:
\be
{\wti g}_{ab} = g_{ab} - \eps \frac{1}{T^2}\pa_a u \pa_b u \quad ,\quad
{\wti g} \equiv \det\Vert {\wti g}_{ab}\Vert \; ,
\lab{ind-metric-ext}
\ee
and ${\wti g}^{ab}$ is the inverse matrix w.r.t. ${\wti g}_{ab}$.

From the definition \rf{ind-metric-ext} and second Eq.\rf{u-T-eqs-NG} one easily finds
that the induced metric on the world-volume is singular on-shell (cf. 
Eq.\rf{on-shell-singular} above):
\be
g_{ab} \( {\wti g}^{bc}\pa_c u\) = 0
\lab{on-shell-singular-A}
\ee
exhibiting the lightlike nature of the $p$-brane described by \rf{LL-action-NG-A}.
Also, comparing \rf{ind-metric-ext} with \rf{gamma-eqs-u} and assuming $\eps=1$
yields relations with the world-volume metric and the dynamical tension from the 
Polyakov-type formulation:
\be
\g_{ab} = \frac{1}{b_0} {\wti g}_{ab} \quad , \quad 
\chi^2 = b_0^{p-1} T^2 \; ,
\lab{Pol-NG-rel}
\ee
with $b_0$ as in \rf{gamma-eqs-u}. Yet the meaning of the constant $b_0$,
which appears as an integration constant within the Polyakov-type formulation
(cf. \rf{gamma-eqs-u}), is different within the Nambu-Goto-type formulation
\rf{LL-action-NG-A} where it 
arises as an arbitrary gauge-fixing parameter for the world-volume reparametrization
invariance -- see \rf{gauge-fix-A} below.

\textbf{Remark 2.} Similarly to the ordinary bosonic $p$-brane we can
rewrite the Nambu-Goto-type action for the \textsl{LL-brane} \rf{LL-action-NG-A} in a
Polyakov-like form by employing an intrinsic Riemannian world-volume metric
$\g_{ab}$ as in \rf{LL-action}:
\be
S_{\rm NG-Pol} = - \h \int d^{p+1}\s\, T\sqrt{-\g}\llb \g^{ab} \( g_{ab} 
- \eps\frac{1}{T^2} \pa_a u \pa_b u\) - \eps b_0 (p-1)\rrb
\lab{LL-action-dual}
\ee
which exhibits more explicitly the duality w.r.t. \rf{LL-action}.

\textbf{Remark 3.} Let us note that duality between the Polyakov-type \rf{LL-action} 
and Nambu-Goto-type \rf{LL-action-NG-A} formulations
of \textsl{LL-brane} dynamic strictly exists provided we choose the sign $\eps=1$
in the Nambu-Goto-type action \rf{LL-action-NG-A}. In what follows we will
use the Nambu-Goto-type formulation to describe one of the \textsl{LL-branes} 
self-consistently coupled to the bulk gravity/gauge-field system where consistency 
of the pertinent solution requires $\eps=-1$.

%%%%%%%%%%%%%%%%%%%%%%%%%%%%%%%%%%%%%%%%%%%%%%%%%%%%%%%%%%%%%%%%%%
\subsection{Coupling to Bulk Space-Time Gauge Fields}
\label{sec:2c}

Using the above world-volume Lagrangian framework one can add in a
natural way \ct{LL-brane-main,inflation-all,our-WH}
couplings of the \textsl{LL-brane} to bulk space-time Maxwell $\cA_\m$ 
and Kalb-Ramond $\cA_{\m_1\ldots\m_{D-1}}$ gauge fields (the latter -- in the case of 
codimension one \textsl{LL-branes}, \textsl{i.e.}, for $D=(p+1)+1$). In the
Polyakov-type formalism we have:
\br
{\wti S}_{\mathrm{Pol}}\lb q,\b\rb = S_{\mathrm{Pol}}
- q \int d^{p+1}\s\,\vareps^{ab_1\ldots b_p} F_{b_1\ldots b_p} \pa_a X^\m \cA_\m
\nonu \\
- \frac{\b}{(p+1)!} \int d^{p+1}\s\,\vareps^{a_1\ldots a_{p+1}}
\pa_{a_1} X^{\m_1}\ldots\pa_{a_{p+1}} X^{\m_{p+1}} \cA_{\m_1\ldots\m_{p+1}}
\lab{LL-action+EM+KR}
\er
with $S_{\mathrm{LL}}$ as in \rf{LL-action}. The \textsl{LL-brane}
constraint equations \rf{phi-eqs}--\rf{gamma-eqs} are not affected by the bulk 
space-time gauge field couplings, whereas Eqs.\rf{A-eqs-0}--\rf{X-eqs-0} acquire the form:
\br
\pa_{[a}\( F^{\ast}_{b]}\, \chi L^\pr (F^2)\) 
+ \frac{q}{4}\pa_a X^\m \pa_b X^\n \cF_{\m\n} = 0  \; ;% \phantom{aaaaa}
\lab{A-eqs} \\
\pa_a \(\chi \sqrt{-\g} \g^{ab} \pa_b X^\m\) + 
\chi \sqrt{-\g} \g^{ab} \pa_a X^\n \pa_b X^\l \G^\m_{\n\l}
% \nonu \\
-q \vareps^{ab_1\ldots b_p} F_{b_1\ldots b_p} \pa_a X^\n \cF_{\l\n}G^{\l\m}
\nonu \\
- \frac{\b}{(p+1)!} \vareps^{a_1\ldots a_{p+1}} \pa_{a_1} X^{\m_1} \ldots
\pa_{a_{p+1}} X^{\m_{p+1}} \cF_{\l\m_1\dots\m_{p+1}} G^{\l\m} = 0 \; .
% \phantom{aaaaa}
\lab{X-eqs}
\er
Here $\chi$ is the dynamical brane tension as in \rf{LL-action-chi}, and
\be
\cF_{\m\n} = \pa_\m \cA_\n - \pa_\n \cA_\m \quad ,\quad
\cF_{\m_1\ldots\m_D} = D\pa_{[\m_1} \cA_{\m_2\ldots\m_D]} =
\cF \sqrt{-G} \vareps_{\m_1\ldots\m_D}
\lab{F-KR}
\ee
are the field-strengths of the electromagnetic $\cA_\m$ and Kalb-Ramond 
$\cA_{\m_1\ldots\m_{D-1}}$ gauge potentials \ct{aurilia-townsend}.
% and:
% \be
% \G^\m_{\n\l}=\h G^{\m\k}\(\pa_\n G_{\k\l}+\pa_\l G_{\k\n}-\pa_\k G_{\n\l}\)
% \lab{affine-conn}
% \ee
% is the Christoffel connection for the external space-time metric.

The dual counterpart of the action \rf{LL-action+EM+KR} within the Nambu-Goto-type 
formalism reads:
\br
{\wti S}_{\rm NG} \lb{\bar q},{\bar\b}\rb = - \int d^{p+1}\s \, T 
\sqrt{\bgv\, \det\Vert g_{ab} - \eps \frac{1}{T^2}
(\pa_a u + {\bar q}\cA_a)(\pa_b u  + {\bar q}\cA_b)\Vert\,\bgv}
% \quad ,\quad \eps = \pm 1 \; .
\nonu \\
- \frac{{\bar\b}}{(p+1)!} \int d^{p+1}\s\,\vareps^{a_1\ldots a_{p+1}}
\pa_{a_1} X^{\m_1}\ldots\pa_{a_{p+1}} X^{\m_{p+1}} \cA_{\m_1\ldots\m_{p+1}}
% \quad , \quad \cA_a \equiv \pa_a X^\m \cA_\m \; .
\lab{LL-action-NG-A+EM+KR}
\er
with $g_{ab}$ denoting the induced metric on the world-volume \rf{ind-metric}
and $\cA_a \equiv \pa_a X^\m \cA_\m$.
Accordingly, using the short-hand notation generalizing \rf{ind-metric-ext}:
\be
{\bar g}_{ab} \equiv g_{ab} - \eps \frac{1}{T^2}
(\pa_a u + {\bar q}\cA_a)(\pa_b u  + {\bar q}\cA_b) 
\quad , \quad \cA_a \equiv \pa_a X^\m \cA_\m \; ,
\lab{ind-metric-ext-A}
\ee
% with ${\wti g}^{ab}$ indicating its inverse, 
the equations of motion w.r.t. $X^\m$, $u$ and $T$ acquire the form:
\br
\pa_a \( T \sqrt{|{\bar g}|} {\bar g}^{ab}\pa_b X^\m\)
+ T \sqrt{|{\bar g}|} {\bar g}^{ab} \pa_a X^\l \pa_b X^\n \G^\m_{\l\n}
\nonu \\
+ \eps \frac{{\bar q}}{T} \sqrt{|{\bar g}|} {\bar g}^{ab}
\pa_a X^\n (\pa_b u  + {\bar q}\cA_b) \cF^{\l\n}G^{\m\l}
\nonu \\
- \frac{{\bar\b}}{(p+1)!} \vareps^{a_1\ldots a_{p+1}} \pa_{a_1} X^{\m_1} \ldots
\pa_{a_{p+1}} X^{\m_{p+1}} \cF_{\l\m_1\dots\m_{p+1}} G^{\l\m} = 0 \; ,
\lab{X-eqs-NG-A}
\er
\br
\pa_a \(\frac{1}{T} \sqrt{|{\bar g}|} {\bar g}^{ab}(\pa_b u  + {\bar q}\cA_b)\) = 0 \; ,
\lab{u-eqs-NG-A} \\
T^2 + \eps {\bar g}^{ab}(\pa_a u  + {\bar q}\cA_a)(\pa_b u  + {\bar q}\cA_b) = 0 \; .
\lab{T-eq-NG-A}
% \lab{u-T-eqs-NG-A}
\er
The on-shell singularity of the induced metric $g_{ab}$ \rf{ind-metric}, 
\textsl{i.e.}, the lightlike property, now reads (cf. Eq.\rf{on-shell-singular-A}):
\be
g_{ab} \({\bar g}^{bc}(\pa_c u  + {\bar q}\cA_c)\) = 0 \; .
\lab{on-shell-singular-A-A}
\ee

%%%%%%%%%%%%%%%%%%%%%%%%%%%%%%%%%%%%%%%%%%%%%%%%%%%%%%%%%%%%%%%%%%
%%%%%%%%%%%%%%%%%%%%%%%%%%%%%%%%%%%%%%%%%%%%%%%%%%%%%%%%%%%%%%%%%%
\section{Lightlike Brane Dynamics in Various Types of Gravitational Backgrounds}
\label{sec:3}
\subsection{Gauge-Fixed Lightlike Brane Equations of Motion}
\label{sec:3a}
Going back to the Polyakov-type formalism (Sec.\ref{sec:2a}), world-volume 
reparametrization invariance allows us to introduce the standard synchronous 
gauge-fixing conditions:
\be
\g^{0i} = 0 \;\; (i=1,\ldots,p) \; ,\; \g^{00} = -1 \; .
\lab{gauge-fix}
\ee
Also, we will use a natural ansatz for the ``electric'' part of the 
auxiliary world-volume gauge field-strength \rf{p-rank}:
\be
F^{\ast i}= 0 \;\; (i=1,{\ldots},p) \quad ,\quad \mathrm{i.e.} \;\;
F_{0 i_1 \ldots i_{p-1}} = 0 \; ,
\lab{F-ansatz}
\ee
meaning that we choose the lightlike direction in Eq.\rf{on-shell-singular} 
to coincide with the brane
proper-time direction on the world-volume ($F^{*\, a}\pa_a \sim \pa_\t$).
The Bianchi identity ($\nabla_a F^{\ast\, a}=0$) together with 
\rf{gauge-fix}--\rf{F-ansatz} and the definition for the dual field-strength
in \rf{p-rank} imply:
\be
\pa_\t \g^{(p)} = 0 \quad \mathrm{where}\;\; \g^{(p)} \equiv \det\Vert\g_{ij}\Vert \; .
\lab{gamma-p-0}
\ee

Taking into account \rf{gauge-fix}--\rf{F-ansatz}, Eqs.\rf{gamma-eqs}
acquire the following gauge-fixed form (recall definition of the induced metric
$g_{ab}$ \rf{ind-metric}):
\be
g_{00}\equiv \Xdot^\m\!\! G_{\m\n}\!\! \Xdot^\n = 0 \quad ,\quad g_{0i} = 0 \quad ,\quad
g_{ij} - b_0\, \g_{ij} = 0 \; ,
\lab{gamma-eqs-0}
\ee
where $b_0$ is the same constant as in \rf{gamma-eqs-u}.

We can impose gauge-fixing of reparametrization invariance within the 
Nambu-Goto-type formalism (Sec.\ref{sec:2b}) similar to \rf{gauge-fix} (using notation 
\rf{ind-metric-ext-A}):
\be
{\bar g}_{0i} = 0 \;\; (i=1,\ldots,p) \quad ,\quad 
{\bar g}_{00} = -\eps {\bar b}_0 \; ,\; {\bar b}_0 = \mathrm{const} >0 \; .
\lab{gauge-fix-A}
\ee
Here the constant ${\bar b}_0$ is an arbitrary gauge-fixing parameter which now can be
identified with the arbitrary integration constant $b_0$ constant \rf{gamma-eqs-u}
within the Polyakov-type formalism.
Also, in analogy with the ansatz \rf{F-ansatz} within the Polyakov-type formulation
we will use the ansatz :
\be
\pa_i u + \cA_i = 0 \; ,
\lab{u-ansatz}
\ee
which is consistent for spherically symmetric bulk space-time Maxwell fields $\cA_\m$
and whose physical meaning is that we choose the lightlike direction in 
Eq.\rf{on-shell-singular-A} to coincide with the brane proper-time direction on the 
world-volume.

With \rf{gauge-fix-A}--\rf{u-ansatz} Eq.\rf{u-eqs-NG-A} implies (cf. Eq.\rf{gamma-p-0}
above):
\be
\pa_0 g^{(p)} = 0 \quad \mathrm{where}\;\; g^{(p)}\equiv\det\Vert g_{ij}\Vert \; ,
\lab{g-p-0}
\ee
$g_{ij}$ being the spacelike part of the induced metric \rf{ind-metric}.

Taking into account \rf{gauge-fix-A}--\rf{g-p-0}, the equations of motion
\rf{u-eqs-NG-A} and \rf{T-eq-NG-A} (or, equivalently, 
\rf{on-shell-singular-A-A}) reduce to (cf. Eqs.\rf{gamma-eqs-0}--\rf{X-eqs-0}):
\br
g_{00} \equiv \Xdot^\m\!\! G_{\m\n}\!\! \Xdot^\n = 0 
\quad ,\quad g_{0i} = 0 \quad ,\quad
% g_{0i} \equiv \Xdot^\m\!\! G_{\m\n} \pa_i X^\n = 0 \; ,
% \nonu \\
T^2 = \frac{1}{{\bar b}_0} \(\pa_0 u + \cA_0\)^2 \;\; (\;\mathrm{i.e.}\;
\pa_i T = 0) \; .
\lab{g-eqs-A} 
\er

%%%%%%%%%%%%%%%%%%%%%%%%%%%%%%%%%%%%%%%%%%%%%%%%%%%
\subsection{Spherically Symmetric Backgrounds}
\label{sec:3b}
Here we will be interested in static spherically symmetric solutions 
of Einstein-Maxwell equations (see Eqs.\rf{Einstein-eqs}--\rf{Maxwell-eqs} below).
We will consider the following generic form of static spherically symmetric metric:
\be
ds^2 = - A(\eta) dt^2 + \frac{d\eta^2}{A(\eta)} + 
C(\eta) h_{ij}(\th) d\th^i d\th^j \; ,
\lab{static-spherical}
\ee
or, in Eddington-Finkelstein coordinates \ct{EFM} ($dt = dv-\frac{d\eta}{A(\eta)}$) :
\be
ds^2 = - A(\eta) dv^2 + 2 dv\,d\eta + C(\eta) h_{ij}(\th) d\th^i d\th^j \; .
\lab{EF-metric}
\ee
Here $h_{ij}$ indicates the standard metric on the sphere $S^p$. 
The radial-like coordinate $\eta$ will vary in general from $-\infty$ to $+\infty$. 

We will consider the simplest ansatz for the \textsl{LL-brane} embedding
coordinates:
\be
X^0\equiv v = \t \quad, \quad X^1\equiv \eta = \eta (\t) \quad, \quad 
X^i\equiv \th^i = \s^i \;\; (i=1,\ldots ,p)
\lab{X-embed}
\ee
Now, the \textsl{LL-brane} equations \rf{gamma-eqs-0} together with \rf{gamma-p-0}
yield:
\be
-A(\eta) + 2\etadot = 0 \quad , \quad 
\pa_\t C = \etadot\,\pa_\eta C\bv_{\eta=\eta(\t)} = 0 \; .
\lab{eta-const}
\ee
First, we will consider the case of $C(\eta)$ as non-trivial function of $\eta$ 
(\textsl{i.e.}, proper spherically symmetric space-time). In this case 
Eqs.\rf{eta-const} imply:
\be
\etadot = 0 \; \to \; \eta (\t) = \eta_0 = \mathrm{const} \quad ,\quad 
A(\eta_0) = 0 \; .
\lab{horizon-standard}
\ee
Eq.\rf{horizon-standard} tells us that consistency of \textsl{LL-brane} dynamics in 
a proper spherically symmetric gravitational background of codimension one requires the 
latter to possess a horizon (at some $\eta = \eta_0$), which is automatically occupied 
by the \textsl{LL-brane} (``horizon straddling'' according to the
terminology of Ref.\ct{Barrabes-Israel}). Similar property -- 
``horizon straddling'', has been found also for \textsl{LL-branes} moving in
rotating axially symmetric (Kerr or Kerr-Newman) and rotating cylindrically
symmetric black hole backgrounds \ct{our-WH,rot-WH}. 
% In what follows we will take $\eta_0 = 0$.

With the embedding ansatz \rf{X-embed} and assuming the bulk Maxwell field
to be purely electric static one ($\cF_{0\eta} = \cF_{v\eta} \neq 0$, the
rest being zero; this is the relevant case to be discussed in what follows), 
Eq.\rf{A-eqs} yields the simple relation:
$\pa_i \chi = 0 \;,\; \mathrm{i.e.}\;\; \chi = \chi (\t)$.
% \be
% \pa_i \chi = 0 \quad ,\quad \mathrm{i.e.}\;\; \chi = \chi (\t) \; .
% \lab{chi-eq-0}
% \ee
Further, the only non-trivial contribution of the second order (w.r.t. world-volume
proper time derivative) $X^\m$-equations of motion \rf{X-eqs} arises for $\m=v$, 
where the latter takes the form of an evolution equation for the dynamical tension
$\chi (\t)$:
\be
\pa_\t \chi + \chi\h \Bigl\lb \pa_\eta A + p b_0 \pa_\eta \ln C\Bigr\rb_{\eta=\eta_0}
-q \sqrt{p! F^2} \cF_{v\eta} + \b b_0^{p/2} \cF = 0 
\lab{chi-eq}
\ee
(recall \rf{L-eq} and the definition \rf{F-KR} of the Kalb-Ramond field-strength $\cF$).
In the case of absence of couplings to bulk space-time gauge fields, Eq.\rf{chi-eq}
yields exponentional ``inflation''/``deflation'' at large times for the dynamical 
\textsl{LL-brane} tension:
\be
\chi (\t) = \chi_0 
\exp\Bigl\{-\t \h\Bigl(\pa_\eta A + p b_0 \pa_\eta \ln C \Bigr)_{\eta=\eta_0}\Bigr\}
\quad ,\quad \chi_0 = \mathrm{const} \; .
\lab{chi-eq-standard-sol}
\ee
Similarly to the ``horizon straddling'' property, exponential 
``inflation''/``deflation'' for the \textsl{LL-brane} tension has also been
found in the case of test \textsl{LL-brane} motion in rotating axially symmetric
and rotating cylindrically symmetric black hole backgrounds 
(for details we refer to Refs.\ct{inflation-all,our-WH,rot-WH}).
This phenomenon is an analog of the ``mass inflation'' effect around black hole
horizons \ct{israel-poisson}.

%%%%%%%%%%%%%%%%%%%%%%%%%%%%%%%%%%%%%%%%%%%%%%%%%%%
\subsection{Product-Type Gravitational Backgrounds: Bertotti-Robinson Space-Time}
\label{sec:3c}
Consider now the case $C(\eta) = \mathrm{const}$ in \rf{EF-metric},
\textsl{i.e.}, the
corresponding space-time manifold is of product type $\S_2 \times S^p$.
% with -- metric on $\S_2$. 
A physically relevant example is the Bertotti-Robinson \ct{BR,lapedes-78}
space-time in $D=4$ (\textsl{i.e.}, $p=2$) with (non-extremal) metric 
in standard coordinates $({\wti t},\z,\th^i)$ (cf.\ct{lapedes-78}): 
\be
ds^2 = r_0^2 \llb - \sinh^2\!\!\z d{\wti t}^2 + d\z^2 + 
d\th^2 + \sin^2 \th d\vp^2 \rrb
% h_{ij}(\th) d\th^i d\th^j \rrb \; ,
\lab{BR-standard}
\ee
describing Anti-de-Sitter${}_2 \times S^2$. 
Upon coordinate change $({\wti t},\z) \to (t,\eta)$:
\be
t = e^{\wti{t}} \coth \z \quad, \quad \eta = e^{-\wti{t}} \sinh \z
\lab{BR-coord-transf}
\ee
the metric \rf{BR-standard} acquires the form:
\be
ds^2 = r_0^2 \llb -\eta^2 dt^2 + \frac{d\eta^2}{\eta^2} + 
% h_{ij}(\th) d\th^i d\th^j \rrb \; ,
d\th^2 + \sin^2 \th d\vp^2 \rrb \; ,
\lab{BR-new}
\ee
or in Eddington-Finkelstein (EF) form ($dt = \frac{1}{r_0^2}dv - \frac{d\eta}{\eta^2}$):
\be
ds^2 = -\frac{\eta^2}{r_0^2}dv^2 + 2 dv d\eta + % r_0^2 h_{ij}(\th) d\th^i d\th^j \, .
r_0^2 \llb d\th^2 + \sin^2 \th d\vp^2 \rrb \; .
\lab{BR-EF}
\ee
At $\eta = 0$ the Bertotti-Robinson metric \rf{BR-new} (or \rf{BR-EF}) possesses a 
horizon. Further, we will consider the case of Bertotti-Robinson universe with
constant electric field $\cF_{v\eta} = \pm \frac{1}{2 r_0 \sqrt{\pi}}$.
% \be
% \cF_{v\eta} = \pm \frac{1}{2 r_0 \sqrt{\pi}} \; .
% \lab{BR-Maxwell}
% \ee
In the present case the second Eq.\rf{eta-const} is trivially satisfied whereas the
first one yields:
$\eta (\t) = \eta (0) \Bigl(1 - \t\frac{\eta (0)}{2 r_0^2}\Bigr)^{-1}$.
% \be
% \eta (\t) = \frac{\eta (0)}{1 - \t\frac{\eta (0)}{2 r_0^2}} \; .
% \lab{BR-eta}
% \ee
In particular, if the \textsl{LL-brane} is initially (at $\t=0$) located on the 
Bertotti-Robinson horizon $\eta = 0$, it will stay there permanently.

%%%%%%%%%%%%%%%%%%%%%%%%%%%%%%%%%%%%%%%%%%%%%%%%%%%%%%%%%%%%%
\section{Lagrangian Formulation of Bulk Gravity/Gauge-Field System Interacting
With Lightlike Branes}
\label{sec:4}
Let us now consider self-consistent bulk Einstein-Maxwell-Kalb-Ramond system coupled 
to {\em two} distantly separated charged codimension-one {\em lightlike}
$p$-branes (in this case $D=(p+1)+1$). It is described by the following action:
\be
S = \int\!\! d^D x\,\sqrt{-G}\,\llb \frac{R(G)}{16\pi} 
- \frac{1}{4} \cF_{\m\n}\cF^{\m\n} 
- \frac{1}{D! 2} \cF_{\m_1\ldots\m_D}\cF^{\m_1\ldots\m_D}\rrb 
+ {\wti S}_{\mathrm{Pol}}\lb q,\b\rb + 
{\wti S}_{\mathrm{NG}}\lb{\bar q},{\bar \b}\rb\; .
\lab{E-M-KR+2LL}
\ee
Here again $\cF_{\m\n}$ and $\cF_{\m_1\ldots\m_D}$ are the Maxwell and Kalb-Ramond
field-strengths \rf{F-KR}. The last two terms on the r.h.s. of \rf{E-M-KR+2LL} 
denote the reparametrization invariant world-volume actions of the two
\textsl{LL-branes} coupled to the bulk space-time gauge fields -- the first
one in the Polyakov-type form \rf{LL-action+EM+KR} and the second one in the
Nambu-Goto-type form \rf{LL-action-NG-A+EM+KR} (below we will need to choose the sign 
$\eps=-1$ in the latter). The objects pertaining to the second \textsl{LL-brane}
will be denoted with bars.

The corresponding Einstein-Maxwell-Kalb-Ramond equations of motion derived from
the action \rf{E-M-KR+2LL} read:
\be
R_{\m\n} - \h G_{\m\n} R =
8\pi \( T^{(EM)}_{\m\n} + T^{(KR)}_{\m\n} + T^{(brane)}_{\m\n} + 
{\bar T}^{(brane)}_{\m\n}\) \; ,
\lab{Einstein-eqs}
\ee
\br
\pa_\n \(\sqrt{-G}\cF^{\m\n}\) + 
q \int\!\! d^{p+1}\s\,\d^{(D)}\Bigl(x-X(\s)\Bigr)
\vareps^{ab_1\ldots b_p} F_{b_1\ldots b_p} \pa_a X^\m 
\nonu \\
-\eps {\bar q} \int\!\! d^{p+1}\s\,\d^{(D)}\Bigl(x-{\bar X}(\s)\Bigr)
\sqrt{|{\bar g}|} {\bar g}^{ab}\pa_a {\bar X}^\m \frac{\pa_b u + {\bar q}\cA_b}{T}
= 0 \; ,
\lab{Maxwell-eqs}
\er
\br
\vareps^{\n\m_1\ldots\m_{p+1}} \pa_\n \cF
- \b\,\int\! d^{p+1}\s\,\d^{(D)}(x - X(\s))
\vareps^{a_1\ldots a_{p+1}} \pa_{a_1}X^{\m_1}\ldots\pa_{a_{p+1}}X^{\m_{p+1}} 
\nonu \\
- {\bar \b}\,\int\! d^{p+1}\s\,\d^{(D)}(x - {\bar X}(\s))
\vareps^{a_1\ldots a_{p+1}} \pa_{a_1}{\bar X}^{\m_1}\ldots\pa_{a_{p+1}}{\bar X}^{\m_{p+1}} 
= 0 \; .
\lab{F-KR-eqs}
\er
Here $X(\s)$ and ${\bar X}(\s)$ denoted the space-time embeddings of the two 
\textsl{LL-branes}, ${\bar g}_{ab}$ is as defined in \rf{ind-metric-ext-A}, 
and in \rf{F-KR-eqs} we have used the last relation \rf{F-KR}. 
The explicit form of the energy-momentum tensors read:
\be
T^{(EM)}_{\m\n} = \cF_{\m\k}\cF_{\n\l} G^{\k\l} - G_{\m\n}\frac{1}{4}
\cF_{\r\k}\cF_{\s\l} G^{\r\s}G^{\k\l} \; ,
\lab{T-EM}
\ee
\br
T^{(KR)}_{\m\n} = \frac{1}{(D-1)!}\llb \cF_{\m\l_1\ldots\l_{D-1}}
{\cF_{\n}}^{\l_1\ldots\l_{D-1}} -
\frac{1}{2D} G_{\m\n} \cF_{\l_1\ldots\l_D} \cF^{\l_1\ldots\l_D}\rrb
\nonu \\
= - \h \cF^2 G_{\m\n}  \; ,
\lab{T-KR}
\er
\be
T_{(brane)}^{\m\n} = 
-\int\!\! d^{p+1}\s\,\frac{\d^{(D)}\Bigl(x-X(\s)\Bigr)}{\sqrt{-G}}\,
\chi\,\sqrt{-\g} \g^{ab}\pa_a X^\m \pa_b X^\n \; ,
\lab{T-brane}
\ee
\be
{\bar T}_{(brane)}^{\m\n} = 
- \int\!\! d^{p+1}\s\,\frac{\d^{(D)}\Bigl(x-{\bar X}(\s)\Bigr)}{\sqrt{-G}}\,
T\,\sqrt{|{\bar g}|} {\bar g}^{ab}\pa_a {\bar X}^\m \pa_b {\bar X}^\n \; .
\lab{T-brane-A}
\ee
The \textsl{LL-brane} stress-energy tensors \rf{T-brane}--\rf{T-brane-A} are 
straightforwardly derived upon varying w.r.t. $G^{\m\n}$ from the world-volume 
actions \rf{LL-action+EM+KR}
% \rf{LL-action} (or, equivalently, \rf{LL-action-chi}) 
and \rf{LL-action-NG-A+EM+KR}, respectively (recall $\chi\equiv\frac{\P}{\sqrt{-\g}}$
is the variable brane tension as in \rf{LL-action-chi}).

The equations of motion for the \textsl{LL-branes} have already been written
down in Sec. \ref{sec:2} -- Eqs.\rf{phi-eqs}--\rf{gamma-eqs} and \rf{A-eqs}--\rf{A-eqs}
for the first \textsl{LL-brane} (in the Polyakov-type formulation) and
Eqs.\rf{X-eqs-NG-A}--\rf{T-eq-NG-A} for the second \textsl{LL-brane} (in the 
Nambu-Goto-type formulation).

In what follows we will employ the embedding ansatz \rf{X-embed} for both 
\textsl{LL-branes} together with \rf{horizon-standard} as well as the world-volume
reparametrization gauge-fixings \rf{gauge-fix} and \rf{gauge-fix-A}.
% \rf{gauge-fix}--\rf{gamma-eqs-0} and , 
In this case the Kalb-Ramond equations of motion \rf{F-KR-eqs} reduce to:
\be
\pa_\eta \cF + \b \d (\eta-\eta_0) + {\bar\b} \d (\eta-{\bar\eta}_0) = 0 
\lab{F-KR-0}
\ee
implying:
\be
\cF\bv_{\eta \to \eta_0 -0} - \cF\bv_{\eta \to \eta_0 +0} = \b \quad ,\quad 
\cF\bv_{\eta \to {\bar\eta}_0 -0} - \cF\bv_{\eta \to {\bar\eta}_0 +0} = {\bar\b}
% \quad ,\quad 
% \cF\bv_{\eta \to \eta_0 +0}=\cF\bv_{\eta \to {\bar\eta}_0 -0}
\lab{F-jump}
\ee
% \cF\bv_{\eta \to \eta_0 \pm 0} = \mathrm{const} \quad ,\quad 
% \cF\bv_{\eta \to {\bar\eta}_0 \pm 0} = \mathrm{const} 
% \er
with $\cF\bv_{\eta \to \eta_0 +0}=\cF\bv_{\eta \to {\bar\eta}_0 -0}$.
Here $\eta_0$ and ${\bar\eta}_0$ indicate the positions of both \textsl{LL-branes}
which according to \rf{horizon-standard} must represent horizons w.r.t. the embedding
space-time geometry. We will assume $\eta_0 < {\bar\eta}_0$ (see below).
Thus, according to Eqs.\rf{F-KR-0}--\rf{F-jump} a piece-wise varying
non-negative cosmological constant $\L = 4\pi \cF^2$ (cf. Eq.\rf{T-KR}) is dynamically 
generated  in the various space-time regions w.r.t. the horizons at $\eta=\eta_0$ and 
$\eta={\bar\eta}_0$, respectively.
% \be
% \L_{(\pm)} = 4\pi \cF^2_{(\pm)} \; .
% \lab{cosmolog-const}
% \ee

Next, we consider static spherically symmetric solution of Maxwell equations
\rf{Maxwell-eqs} which in this case, using ansatz \rf{F-ansatz} and the 
gauge-fixing conditions \rf{gauge-fix-A} together with \textsl{LL-brane} embeddings 
\rf{X-embed}, acquire the form:
\be
\pa_\eta \( C^{p/2} (\eta) \cF_{v\eta} (\eta)\)
- q\,\frac{\sqrt{p! F^2}}{b_0^{p/2}} C^{p/2}(\eta_0) \d (\eta - \eta_0)
-{\bar q} C^{p/2}({\bar\eta}_0) \d (\eta - {\bar\eta}_0) = 0 \; .
\lab{Maxwell-0}
\ee

As mentioned above in Sec. \ref{sec:3}, the evolution equation \rf{chi-eq} for the 
dynamical \textsl{LL-brane} tension $\chi$ remains the only non-trivial contribution 
of the second order $X^\m$-equations of motion \rf{X-eqs} for $\m = v$ 
within the Polyakov-type formulation when employing the embedding ansatz \rf{X-embed}. 
% Similar equation exists
% for the dynamical \textsl{LL-brane} tension $T$ within the dual Nambu-Goto-type 
% formulation. 
In the present context of looking for self-consistent solution of the
bulk gravity/gauge-field system coupled to the two \textsl{LL-branes} \rf{E-M-KR+2LL}
the ``force'' terms in \rf{chi-eq} (the geodesic ones containing the Christoffel 
connection coefficients as well as those coming from the \textsl{LL-brane} coupling to 
the bulk Maxwell and Kalb-Ramond gauge fields) contain discontinuities across the 
horizon occupied by the corresponding \textsl{LL-brane}. The discontinuity problem 
is resolved following the approach in Ref.\ct{Israel-66} (see also the regularization
approach in Ref.\ct{BGG}, Appendix A) by taking mean values of the ``force''
terms across the discontinuity at $\eta=\eta_0$. Thus, Eq.\rf{chi-eq} is replaced by:
% we obtain from Eq.\rf{X-eqs} with $\m=v$:
\be
\pa_\t \chi + \chi\h \( \llangle \pa_\eta A \rrangle_{\eta=\eta_0} 
+ p b_0 \llangle \pa_\eta \ln C \rrangle_{\eta=\eta_0} \)
-q \sqrt{p! F^2} \llangle \cF_{v\eta}\rrangle_{\eta=\eta_0} 
+ \b b_0^{p/2} \llangle \cF\rrangle_{\eta=\eta_0} = 0 
\lab{chi-eq-mean}
\ee
Here and below we will use the following short-hand notations for ``mean
value'' and ``discontinuity'' for any relevant quantity across $\eta=\eta_0$:
\be
\llangle Y \rrangle_{\eta=\eta_0} \equiv 
\h \( Y\bv_{\eta \to \eta_0 +0} + Y\bv_{\eta \to \eta_0 -0}\)  \quad ,\quad
\bigl\lb Y \bigr\rb_{\eta=\eta_0} \equiv 
Y\bv_{\eta \to \eta_0 +0} - Y\bv_{\eta \to \eta_0 -0} \; ,
\lab{mean-discont}
\ee
and similarly across $\eta={\bar\eta}_0$.

Accordingly, the $X^\m$-equation for $\m = v$ within the dual Nambu-Goto-type
formulation \rf{X-eqs-NG-A} taking into account the gauge-fixing \rf{gauge-fix-A} 
becomes evolution equation for the dynamical \textsl{LL-brane} tension $T$
which we will write for convenience in terms of the rescaled quantity:
\be
{\bar\chi} \equiv T {\bar b}_0^{\frac{p-1}{2}}
\lab{chi-bar-def}
\ee
(cf. second relation in \rf{Pol-NG-rel}) :
\be
% \pa_\t T + T\h \Bigl\lb \llangle \pa_\eta A \rrangle_{\eta={\bar\eta}_0} 
% + p {\bar b}_0 \llangle \pa_\eta \ln C \rrangle_{\eta={\bar\eta}_0} \Bigr\rb
% - \eps \sqrt{b_0} {\bar q} \llangle \cF_{v\eta}\rrangle_{\eta={\bar\eta}_0} 
% + \eps \sqrt{b_0} {\bar\b} \llangle \cF\rrangle_{\eta={\bar\eta}_0} = 0
\pa_\t {\bar\chi} + 
{\bar\chi} \h \( \llangle \pa_\eta A \rrangle_{\eta={\bar\eta}_0} 
+ p {\bar b}_0 \llangle \pa_\eta \ln C \rrangle_{\eta={\bar\eta}_0} \)
- \eps {\bar b}_0^{p/2} {\bar q} \llangle \cF_{v\eta}\rrangle_{\eta={\bar\eta}_0} 
+ \eps {\bar b}_0^{p/2} {\bar\b} \llangle \cF\rrangle_{\eta={\bar\eta}_0} = 0
\lab{T-eq-mean}
\ee

Taking into account the embedding ansatz \rf{X-embed} for both \textsl{LL-branes} 
together with \rf{horizon-standard}, \rf{gauge-fix} and \rf{gauge-fix-A}, 
the \textsl{LL-brane} energy-momentum tensors \rf{T-brane}--\rf{T-brane-A} on the 
r.h.s. of the Einstein equations of motion \rf{Einstein-eqs}, which were
derived from the underlying world-volume \textsl{LL-brane} actions \rf{LL-action-chi} 
and \rf{LL-action-NG-A+EM+KR}, acquire the form:
\be
T_{(brane)}^{\m\n} = S^{\m\n}\,\d (\eta-\eta_0) \quad ,\quad
{\bar T}_{(brane)}^{\m\n} = {\bar S}^{\m\n}\,\d (\eta-{\bar\eta}_0) 
\lab{T-S-0}
\ee
with surface energy-momentum tensors:
\br
S^{\m\n} \equiv \frac{\chi}{b_0^{p/2}}\,
\( \pa_\t X^\m \pa_\t X^\n - b_0 G^{ij} \pa_i X^\m \pa_j X^\n 
\)_{v=\t,\,\eta=\eta_0,\,\th^i =\s^i} \; ,
\lab{T-S-brane} \\
{\bar S}^{\m\n} \equiv \frac{{\bar\chi}}{\eps {\bar b}_0^{p/2}}\,
\( \pa_\t {\bar X}^\m \pa_\t {\bar X}^\n
- \eps {\bar b}_0 G^{ij} \pa_i {\bar X}^\m \pa_j {\bar X}^\n 
\)_{v=\t,\,\eta={\bar\eta}_0,\,\th^i =\s^i} \; .
\lab{T-S-brane-A}
\er
Here and below $G_{ij} = C(\eta) h_{ij}(\th)$ (cf. \rf{EF-metric}) and
${\bar\chi}$ is as in \rf{chi-bar-def}. For the non-zero components of 
\rf{T-S-brane}--\rf{T-S-brane-A} (with lower indices) and its trace we find:
\br
S_{\eta\eta} = \frac{\chi}{b_0^{p/2}} \quad ,\quad 
S_{ij} = - \frac{\chi}{b_0^{p/2-1}} G_{ij} \quad ,\quad
S^\l_\l = - \frac{p\chi}{b_0^{p/2-1}} 
\lab{S-comp} \\
{\bar S}_{\eta\eta} = \eps\frac{{\bar\chi}}{{\bar b}_0^{p/2}} \quad ,\quad 
{\bar S}_{ij} = - \frac{{\bar\chi}}{{\bar b}_0^{p/2-1}} G_{ij} \quad ,\quad
{\bar S}^\l_\l = - \frac{p{\bar\chi}}{{\bar b}_0^{p/2-1}} 
\lab{S-comp-A}
\er

Because of the delta-function nature of the \textsl{LL-brane} stress-energy
tensors \rf{T-S-0}, the solution of the Einstein-Maxwell-Kalb-Rammond equations 
\rf{Einstein-eqs}--\rf{F-KR-eqs} consists of 
finding ``electrovacuum'' solutions in each of the three space-time regions 
($\eta <\eta_0$), ($\eta_0 < \eta < {\bar\eta}_0$) and ($\eta >
{\bar\eta}_0$) and subsequently matching them along the lightlike
hypersurfaces ($\eta = \eta_0$) and ($\eta = {\bar\eta}_0$) which themselves must be
their pairwise common horizons and which due to \rf{horizon-standard} will be 
automatically occupied by the two \textsl{LL-branes} (``horizon straddling'').
% the \textsl{LL-brane} world-volumes ($\eta = \eta_0$) and ($\eta = {\bar\eta}_0$).

The systematic formalism for matching different bulk space-time geometries on
codimension-one timelike hypersurfaces (``thin shells'') was developed 
originally in Ref.\ct{Israel-66} and later generalized in Ref.\ct{Barrabes-Israel} 
to the case of lightlike hypersurfaces (``null thin shells'') (for a
systematic introduction, see the textbook \ct{poisson-kit}).
In the present case we are interested in static spherically symmetric solutions
with metric of the form \rf{EF-metric}, therefore, due to the simple geometry
% (spherical symmetry and matching on common horizon) 
one can straightforwardly isolate the terms from the Ricci tensor on the l.h.s. of 
Einstein equations \rf{Einstein-eqs} which may yield delta-function contributions 
($\sim \d (\eta - \eta_0)$ and $\sim \d (\eta - {\bar\eta}_0)$) to be matched
with the components of the \textsl{LL-brane} surface stress-energy tensors
\rf{T-S-0}--\rf{T-S-brane-A}. The metric \rf{EF-metric} is continuous at both 
lightlike hypersurfaces ($\eta = \eta_0$) and ($\eta = {\bar\eta}_0$), 
but its first derivative w.r.t. $\eta$ (the normal coordinate w.r.t. both horizons)
might exhibit discontinuity 
across $\eta = \eta_0$ and $\eta = {\bar\eta}_0$. Thus, the terms contributing to 
delta-function singularities in $R_{\m\n}$ are those containing second derivatives
w.r.t. $\eta$. Separating explicitly
the latter we can rewrite Eqs.\rf{Einstein-eqs} in the following form:
% (here we take $D=4$ and $p=2$):
\br
R_{\m\n} \equiv \pa_\eta \G^\eta_{\m\n} - \pa_\m \pa_\n \ln \sqrt{-G}
+ \mathrm{non-singular ~terms}
\nonu \\
= 8\pi \( S_{\m\n} - \frac{1}{p} G_{\m\n} S^{\l}_{\l}\) \d (\eta - \eta_0) +
\nonu \\
8\pi \( {\bar S}_{\m\n} - \frac{1}{p} G_{\m\n} {\bar S}^{\l}_{\l}\) 
\d (\eta - {\bar\eta}_0) + \mathrm{non-singular ~terms} \; .
\lab{Einstein-eqs-sing}
\er

%%%%%%%%%%%%%%%%%%%%%%%%%%%%%%%%%%%%%%%%%%%%%%%%%%%
For the sake of simplicity we will consider in what follows the case of 
$D=4$-dimensional bulk space-time and, correspondingly, $p=2$ for the 
\textsl{LL-branes}. The generalization to arbitrary $D$ is straightforward.
For further simplification of the numerical constant factors we will choose
the following specific (``wrong-sign'' Maxwell) form for the Lagrangian of the 
auxiliary non-dynamical world-volume gauge field in the Polyakov-type \textsl{LL-brane}
formulation (cf. Eqs.\rf{p-rank}--\rf{F2-id}): 
\be
L(F^2)=\frac{1}{4}F^2 \quad \to \quad  b_0 = 2M \; ,
\lab{L-eq-0}
\ee
where again $b_0$ is the constant defined in \rf{gamma-eqs-u} and
$M$ denotes the original integration constant in Eqs.\rf{phi-eqs}. 
%%%%%%%%%%%%%%%%%%%%%%%%%%%%%%%%%%%%%%%%%%%%%%%%%%%

Substituting in Einstein equations \rf{Einstein-eqs-sing} the explicit form of the
metric \rf{EF-metric} and the \textsl{LL-brane} stress-energy expressions 
\rf{S-comp}--\rf{S-comp-A} yields the following matching relations at the 
delta-function singularities(using notations \rf{mean-discont} and \rf{chi-bar-def}):
\br
\llb \pa_\eta A \rrb_{\eta_0} = - 16\pi \chi \quad,\quad
\llb \pa_\eta A \rrb_{{\bar\eta}_0} = - 16\pi {\bar\chi}
\lab{Einstein-eqs-0} \\
\llb \pa_\eta \ln C \rrb_{\eta_0} = - \frac{8\pi}{b_0} \chi  \quad,\quad
\llb \pa_\eta \ln C \rrb_{{\bar\eta}_0} = - \eps\frac{8\pi}{{\bar b}_0}{\bar\chi}
\lab{Einstein-eqs-1}
\er
Consistency of \rf{Einstein-eqs-0}--\rf{Einstein-eqs-1} requires that the
dynamical tensions $\chi,\,{\bar\chi}$ of both \textsl{LL-branes} must be 
time-independent, \textsl{i.e.}, $\pa_\t\chi=0$ and $\pa_\t{\bar\chi}=0$ in
\rf{chi-eq-mean} and \rf{T-eq-mean}, respectively.

From the Maxwell equations \rf{Maxwell-0} we obtain the following matchings for 
the static spherically symmetric electromagnetic field-strength across the 
\textsl{LL-brane} world-volume hypersurfaces (using again notations 
\rf{mean-discont}):
\be
\llb \cF_{v\eta} \rrb_{\eta_0} = \frac{2}{\sqrt{b_0}}\, q  \quad,\quad
\llb \cF_{v\eta} \rrb_{{\bar\eta}_0} = {\bar q}
\lab{Maxwell-eqs-0}
\ee

%%%%%%%%%%%%%%%%%%%%%%%%%%%%%%%%%%%%%%%%%%%%%%%%%%%
\section{Two-Throat Wormhole-Like Solution via Lightlike Branes}
\label{sec:5}
We will seek self-consistent solution of the equations of motion of the
coupled Einstein-Maxwell-Kalb-Ramond-\textsl{LL-brane} system (Section \ref{sec:4})
describing an asymmetric wormhole-like space-time with spherically
symmetric geometry and two ``throats''. 
The general form of asymmetric wormhole-like metric (in Eddington-Finkelstein
coordinates) \rf{EF-metric} specifically reads (now $D=4\,,\,p=2$):
\br
ds^2 = - A(\eta) dv^2 + 2 dv d\eta + 
C(\eta) \llb d\th^2 + \sin^2 \th d\vp^2 \rrb \, ,
\lab{asymm-WH-EF} \\
A (\eta_0) = 0 \quad ,\quad A ({\bar\eta}_0) = 0 \quad ,\quad
A(\eta) > 0 \;\; \mathrm{for ~all}\; \eta \neq \eta_0,\,{\bar\eta}_0 \;\; ,\;\;
{\bar\eta}_0 > \eta_0 \; .
\lab{asymm-WH-A}
\er
% Furthermore, we will require $\chi = \mathrm{const}$ and ${\bar\chi} = \mathrm{const}$
% (independent of the \textsl{LL-brane} proper times) for consistency with the 
% matching relations \rf{Einstein-eqs-0}--\rf{Einstein-eqs-1}. 

The present wormhole-like solution describes {\em three} pairwise matched space-time 
regions:

(i) ``left'' Bertotti-Robinson ``universe'' (cf.\rf{BR-EF}) where:
\be
A(\eta) = \frac{\eta^2}{r_0^2} \quad, \quad C(\eta) = r_0^2 
% \quad, \quad \cF_{v\eta}\bv_{BR} = \pm \frac{1}{2\sqrt{\pi}\,r_0}
\quad, \quad \cF_{v\eta} = \vareps_L \frac{1}{2\sqrt{\pi}\,r_0}
\quad \mathrm{for}\; \eta < 0 \;\; ,\;\; \vareps_L = \pm 1 \; ,
\lab{left-BR}
\ee
with the choice $\eta_0 = 0$ for simplicity;

(ii) ``middle'' Reissner-Nordstr{\"o}m-de-Sitter ``universe'' with:
% $\( A(\eta) \equiv A_{\mathrm{RN}}(r_0 + \eta)\; ,\; 
% \cF_{v\eta}(\eta) \equiv \cF_{0r}\bv_{RN} (r_0 + \eta)\)$ :
\br
A(\eta) \equiv A_{\mathrm{RN}}(r_0 + \eta) = 
1 - \frac{2m}{r_0 + \eta} + \frac{Q^2}{(r_0 + \eta)^2} - 
\frac{4\pi\b^2}{3} (r_0 + \eta)^2 \; ,
\lab{mid-RN-0} \\
C(\eta) = (r_0 + \eta)^2 \;\; ,\;\;
\cF_{v\eta} \equiv \cF_{vr}\bv_{RN} = \frac{Q}{\sqrt{4\pi} (r_0 + \eta)^2}
\quad \mathrm{for}\; 0 < \eta < {\bar\eta}_0 \equiv {\bar r}_0 - r_0 \;\; ,
\lab{mid-RN-1} \\
A(0) \equiv A_{\mathrm{RN}}(r_0) = 0 \quad ,\quad 
\pa_\eta A \bv_{\eta \to +0} \equiv \pa_r A_{\mathrm{RN}}\bv_{r=r_0} >0 \; ,
\lab{mid-RN-2} \\
A({\bar\eta}_0) \equiv A_{\mathrm{RN}}({\bar r}_0) = 0 \quad ,\quad 
\pa_\eta A \bv_{\eta \to {\bar\eta}_0 -0} \equiv 
\pa_r A_{\mathrm{RN}}\bv_{r={\bar r}_0} <0 \; ,
\lab{mid-RN-3}
\er
where $r_0$ and ${\bar r}_0$ (${\bar r}_0 > r_0$) are the intermediate (outer 
Reissner-Nordstr{\"o}m) and the outmost (de-Sitter) horizons of the standard 
Reissner-Nordstr{\"o}m-de-Sitter space-time, respectively;

(iii) another ``right''  Bertotti-Robinson ``universe'' (cf.\rf{BR-EF}) with:
\br
A(\eta) = \frac{\(\eta - {\bar r}_0 + r_0 \)^2}{{\bar r}_0^2} 
\quad, \quad C(\eta) = {\bar r}_0^2 
\nonu \\
\cF_{v\eta} = \vareps_R \frac{1}{2\sqrt{\pi}\,{\bar r}_0}
\quad \mathrm{for}\; \eta > {\bar\eta}_0 \equiv {\bar r}_0 - r_0 
\;\; ,\;\; \vareps_R = \pm 1 \; .
\lab{right-BR}
\er

Let us stress that the cosmological constant:
\be
\L = 4\pi\b^2
\lab{cosmolog-const}
\ee
in the middle Reissner-Nordstr{\"o}m-de-Sitter space-time region (ii) is dynamically
generated by the presence of the \textsl{LL-branes} according to 
\rf{F-KR-0}--\rf{F-jump}, which in particular implies $\b = - {\bar\b}$.

$\cF_{v\eta}$'s are the respective Maxwell field-strengths and the
Reissner-Nordstr{\"o}m charge parameter $Q$ is determined from the discontinuities
of $\cF_{v\eta}$ in Maxwell equations \rf{Maxwell-0} across each of the two charged
\textsl{LL-branes}:
\be
Q = r_0 \Bigl\lb \vareps_L + \sqrt{\frac{\pi}{b_0}}\, 4q r_0 \Bigr\rb
% \lab{Q-1}
% \ee
% and
% \be
\quad \mathrm{and} \quad
Q = {\bar r}_0 \Bigl\lb \vareps_R - \sqrt{4\pi}{\bar q} {\bar r}_0 \Bigr\rb \; .
\lab{Q-rel}
\ee
In \rf{mid-RN-0}--\rf{mid-RN-1} we have used the standard
coordinate notations for the Reissner-Nordstr{\"o}m-de-Sitter metric coefficients and
Coulomb field strength:
\be
A_{\mathrm{RN}}(r) = 1 - \frac{2m}{r} + \frac{Q^2}{r^2} - \frac{\L}{3}r^2 \quad,\quad 
\cF_{vr}\bv_{RN} = \cF_{0r}\bv_{RN} = \frac{Q}{\sqrt{4\pi} r^2} \; .
\lab{RN-standard}
\ee

It now remains to substitute the expressions \rf{left-BR}--\rf{right-BR} into the 
set of equations \rf{Einstein-eqs-0}--\rf{Einstein-eqs-1} (resulting from the
Einstein equations \rf{Einstein-eqs-sing}), Eqs.\rf{Maxwell-eqs-0} (resulting 
from the Maxwell equations \rf{Maxwell-0}) and the equations for the
dynamical brane tensions \rf{chi-eq-mean} and \rf{T-eq-mean} (with vanishing
time-derivative terms, cf. \rf{Einstein-eqs-0}--\rf{Einstein-eqs-1} above).
This will yield a set of algebraic relations determining all parameters of
the ``two-throat'' wormhole-like solution \rf{left-BR}--\rf{right-BR} as
functions of the three free parameters $\b (=-{\bar\b}),\, q,\, {\bar q}$ -- 
the coupling constants of the two \textsl{LL-branes}.

First, Eqs.\rf{Einstein-eqs-1} yield:
\be
r_0 = \frac{b_0}{4\pi |\chi|} \quad ,\quad 
{\bar r}_0 = \frac{{\bar b}_0}{4\pi |{\bar\chi}|} \; ,
\lab{r-0-eqs}
\ee
where consistency requires:
\be
\chi < 0 \;\; ,\;\; {\bar\chi} < 0 \;\; ,\;\; \eps = -1 \; .
\lab{chi-signs}
\ee
Inserting expressions \rf{r-0-eqs} into \rf{mid-RN-2}--\rf{mid-RN-3} allows
to express the mass parameter $m$ of the ``middle'' Reissner-Nordstr{\"o}m-de-Sitter
``universe'' in two ways:
\be
m = \frac{b_0}{4\pi |\chi|} \Bigl( 1 - 2b_0 -\frac{\b^2 b_0^2}{6\pi\chi^2}\Bigr)
\quad,\quad
m = \frac{{\bar b}_0}{4\pi |{\bar\chi}|} \Bigl( 1 + 2{\bar b}_0 
-\frac{\b^2 {\bar b}_0^2}{6\pi{\bar\chi}^2}\Bigr)
\lab{m-rel}
\ee

Further, Eqs.\rf{r-0-eqs} together with \rf{Einstein-eqs-0}, \rf{chi-eq-mean}, \rf{T-eq-mean}
and \rf{Maxwell-eqs-0} imply quadratic equations for the \textsl{LL-brane}
tensions:
\br
\chi^2 + \frac{\vareps_L q}{\sqrt{4\pi b_0}} |\chi| 
+ \frac{1}{16\pi} \( b_0 \b^2 + 4q^2\) = 0 \; ,
\lab{chi-2-eq} \\
{\bar\chi}^2 + \frac{\vareps_R {\bar q}}{4\sqrt{\pi}} |{\bar\chi}| 
- \frac{{\bar b}_0}{16\pi} \(\b^2 + {\bar q}^2\) = 0 \; ,
\lab{chi-2-eq-A}
\er
%%%%%%%%%%%
In what follows it is more convenient to rescale the charge parameter of the first 
\textsl{LL-brane}:
\be
q \to {\widehat q} \equiv \frac{2}{\sqrt{b_0}} q
\lab{q-rescale}
\ee
so that ${\widehat q}$ has the same physical meaning of \textsl{LL-brane}
surface electric charge density as ${\bar q}$ (cf. Eqs.\rf{Maxwell-0} and 
\rf{Maxwell-eqs-0}). With the notation \rf{q-rescale}
%%%%%%%%%%
Eq.\rf{chi-2-eq} acquires a form completely analogous to \rf{chi-2-eq-A}:
\be
\chi^2 + \frac{\vareps_L {\widehat q}}{4\sqrt{\pi}} |\chi| 
+ \frac{b_0}{16\pi} \(\b^2 + {\widehat q}^2\) = 0 \; .
\lab{chi-2-eq-0}
\ee
Existence of solutions to \rf{chi-2-eq-0} (or \rf{chi-2-eq}) requires the
following condition on the signs of the constant electric field of the
``left'' Bertotti-Robinson ``universe'' \rf{left-BR} and the charge of the ``left''
\textsl{LL-brane}:
\be
\vareps_L \,\mathrm{sign}({\widehat q}) = -1 \; .
\lab{signs-1}
\ee
Thus, the solutions of \rf{chi-2-eq-0} and \rf{chi-2-eq-A} read (using notation 
\rf{q-rescale}):
\br
|\chi| = \frac{|{\widehat q}|}{8\sqrt{\pi}} \Bigl\lb 1 \pm 
\sqrt{1 - 4b_0 (1 + \b^2/{\widehat q}^2 )}\Bigr\rb \;\; ,
\lab{chi-barchi-eq-1} \\
|{\bar\chi}| = \frac{|{\bar q}|}{8\sqrt{\pi}} \Bigl\lb -\vareps_R\,\mathrm{sign}({\bar q}) 
+ \sqrt{1 + 4{\bar b}_0 (1 + \b^2/{\bar q}^2 )}\Bigr\rb \; .
\lab{chi-barchi-eq-2}
\er
The expression \rf{chi-barchi-eq-1} implies a constraint on the parameter $b_0$:
\be
b_0 < 1/4(1 + \b^2/{\widehat q}^2 ) \; . %% \;\; \mathrm{or,~ equivalently}, \;\; 
% b_0 < \frac{2 q^2}{\b^2}\(\sqrt{1 + \b^2/4q^2} -1\)
\lab{b0-constr}
\ee
% in terms of the original $q$ \rf{q-rescale}.
% The expression \rf{chi-barchi-eq-1} implies a constraint 
% $b_0 < 1/4(1 + \b^2/{\widehat q}^2 )$ on the parameter $b_0$, 

Finally, we need to insert the expressions \rf{r-0-eqs} and 
\rf{chi-barchi-eq-1}--\rf{chi-barchi-eq-2} into
\rf{Q-rel} and \rf{m-rel} which yield accordingly (taking into account \rf{signs-1}):
% To this end it is more convenient to use the
% following combinations:
% \be
% \O \equiv \frac{1}{\b^2 + {\widehat q}^2} 
% \Bigl(\frac{|{\widehat q}|}{4\sqrt{\pi}} - |\chi|\Bigr) \quad ,\quad
% {\bar \O} \equiv \frac{1}{\b^2 + {\bar q}^2} 
% \Bigl(\frac{\vareps_R\,|{\bar q}|}{4\sqrt{\pi}} + |{\bar\chi}|\Bigr) \; .
% \lab{Omega-eqs}
% \ee
% With the help of \rf{Omega-eqs}, Eqs.\rf{Q-rel} and \rf{m-rel} yield accordingly:
\br
r_0\,\vareps_L \Bigl\lb 1 -  2\sqrt{\pi} |{\widehat q}|r_0\Bigr\rb =
{\bar r}_0 \,\vareps_R \Bigl\lb 1 - 2\sqrt{\pi}\vareps_R\,{\bar q}{\bar r}_0\Bigr\rb \; ,
\lab{Q-rel-Om} \\
r_0\,\Bigl\lb 1 - 2\sqrt{\pi} |{\widehat q}|r_0 
+ 2\pi ({\widehat q}^2 - \frac{\b^2}{3}) r_0^2\Bigr\rb =
{\bar r}_0\,\Bigl\lb 1 - 2\sqrt{\pi}\vareps_R\,{\bar q}{\bar r}_0 
+ 2\pi ({\bar q}^2 - \frac{\b^2}{3}){\bar r}_0^2\Bigr\rb \; .
\lab{m-rel-Om}
\er
The consistent solutions of \rf{Q-rel-Om}--\rf{m-rel-Om} read:
\be
{\bar r}_0 = \a r_0 \quad ,\quad r_0 = 
\frac{\a -1}{2\sqrt{\pi}\(\a^2|{\bar q}| - |{\widehat q}|\)} \; ,
\lab{Omega-sol}
\ee
where:
\be
\a \equiv \Bigl(\frac{{\widehat q}^2 - \b^2/3}{{\bar q}^2 - \b^2/3}\Bigr)^{1/3} >1
\quad ,\quad \a^2 > \frac{|{\widehat q}|}{|{\bar q}|}
\lab{Omega-cond}
\ee
together with the following conditions on the sign factors -- either:
% (here we have reintroduced the original charge parameter $q$, cf.\rf{q-rescale}):
\be
{\widehat q}<0 \;\; ,\;\; {\bar q}>0  \quad ,\quad \vareps_L =\vareps_R = +1
\lab{choice-1}
\ee
or:
\be
{\widehat q}>0 \;\; ,\;\; {\bar q}<0  \quad ,\quad \vareps_L =\vareps_R = -1
\lab{choice-2}
\ee
\textsl{i.e.}, the electric charges of both \textsl{LL-branes} must be opposite 
w.r.t. each other and the constant electric fields in the ``left'' and ``right'' 
Bertotti-Robinson ``universes'' must have equal signs.

Inequalities \rf{Omega-cond} and \rf{b0-constr} imply explicitly the following 
allowed ranges of the free parameters $(\b, {\widehat q}, {\bar q})$ of the 
``two-throat'' wormhole-like solution \rf{left-BR}--\rf{right-BR}:
\be
% \b^2 < 3 {\bar q}^2 \;\; ,\;\; 16 < \frac{{\bar q}^2}{q^2} < 64 \;\;
% (\mathrm{sufficient~condition}) 
% \b^2 > 3 {\bar q}^2 \;\; ,\;\; \frac{{\bar q}^2}{q^2} > 64  \; .
\mathrm{either} \quad \b^2 < 3 {\bar q}^2 \; ,\;\; |{\widehat q}| > |{\bar q}|
\quad \mathrm{or} \quad 
\b^2 > 3 {\bar q}^2 \; ,\;\; |{\widehat q}| < |{\bar q}| \; ,
\lab{inequal-A-B}
\ee
and (using the short-hand notation $\a$ from \rf{Omega-cond}):
\be
\frac{(\a -1)|{\widehat q}| \(1 + \b^2/{\widehat q}^2\)}{\a^2 |{\bar q}| 
- |{\widehat q}|} < 2 \; .
\lab{inequal-C}
\ee
% \be
% |{\widehat q}| > |{\bar q}| \quad ,\quad
% \b^2 < 3 \(|{\widehat q}| |{\bar q}|]\)^{3/2}
% \frac{|{\widehat q}| + |{\widehat q}|^{1/2} |{\bar q}|^{1/2} + 
% |{\bar q}|}{{\widehat q}^2 + |{\widehat q}| |{\bar q}| + {\bar q}^2} \; .
% \lab{choice}
% \ee
Inequalities \rf{inequal-A-B}--\rf{inequal-C} imply upper and lower bounds on 
the dynamically generated (due to the \textsl{LL-branes}) cosmological constant
$\L = 4\pi \b^2$ \rf{cosmolog-const} in the middle Reissner-Nordstr{\"o}m-de-Sitter
space-time region \rf{mid-RN-0}--\rf{mid-RN-3}. In particular, we find that $\L$
cannot be arbitrary small.

%%%%%%%%%%%%%%%%%%%%%%%%%%%%%%%%%%%%%%%%%%%%%%%%%%%%%%%%%%%%%%%%%%%%%
Let us also note that for a special fine-tuning of the \textsl{LL-brane} parameters 
$\a = |{\widehat q}|/|{\bar q}|$ it follows from \rf{Q-rel} and \rf{Omega-sol}
that $Q=0$, \textsl{i.e.}, the electric field inside the ``middle'' universe
\rf{mid-RN-0}--\rf{mid-RN-3} vanishes and the latter becomes an electrically neutral
Schwarzschild-de-Sitter region.

%%%%%%%%%%%%%%%%%%%%%%%%%%%%%%%%%%%%%%%%%%%%%%%%%%%%%%%%%%%%%
\section{Discussions and Conclusions}
\label{sec:6}
% \section{Traversability}

In the present work we have explored the use of (codimension-one) \textsl{LL-branes} for 
construction of new asymmetric wormhole-like solutions of Einstein-Maxwell-Kalb-Ramond
system self-consistently coupled to two widely separated \textsl{LL-branes}
which describe a sequence of spontaneous compactification/decompactification
transitions of space-time. In the course of work we have strongly emphasized the 
crucial properties of the dynamics of \textsl{LL-branes} interacting with gravity and
bulk space-time gauge fields:

(i) ``Horizon straddling'' -- automatic positioning of \textsl{LL-branes} on (one of) 
the horizon(s) of the bulk space-time geometry; 

(ii) Intrinsic nature of the \textsl{LL-brane} tension as an additional 
{\em dynamical degree of freedom} unlike the case of standard Nambu-Goto $p$-branes;
(where it is a given \textsl{ad hoc} constant), and which might in particular acquire
negative values;

(iii) The stress-energy tensors of the \textsl{LL-branes} are systematically derived
from the underlying \textsl{LL-brane} Lagrangian actions and provide the appropriate
source terms on the r.h.s. of Einstein equations to enable the existence of
consistent non-trivial wormhole-like solutions; 

(iv) Electrically charged \textsl{LL-branes} naturally produce {\em asymmetric}
wormholes with the \textsl{LL-branes} themselves materializing the wormhole ``throats''
and uniquely determining the pertinent wormhole parameters.

% % (iv) Electrically neutral \textsl{LL-branes} produce {\em symmetric}
% % wormholes, \textsl{i.e.}, where both left and right ``universes'' are
% % \textsl{i.e.}, where both left and right ``universes'' are
% % related via reflection symmetry and where, in particular, the Misner-Wheeler
% % ``charge without charge'' \ct{misner-wheeler} phenomenon is observed;

(v) \textsl{LL-branes} naturally couple to 3-index Kalb-Ramond bulk
space-time gauge fields which results in {\em dynamical} generation of space-time
varying cosmological constant.

% % (v) The wormhole reflection symmetry is broken through the natural couplings
% % of the \textsl{LL-brane} to bulk space-time gauge fields (Maxwell and
% % 3-index Kalb-Ramond). In this way the \textsl{LL-brane} dynamically generates 
% % non-zero Coulomb field-strength in the ``right'' universe and non-zero cosmological 
% % constant either in the ``left'' or in the ``right'' universe which enable the existence 
% % of {\em asymmetric} (with {\em no} reflection symmetry) wormholes.

Let us point out that the above asymmetric ``two-throat'' wormhole, connecting
through the first ``throat'' a ``left'' compactified Bertotti-Robinson universe to a 
``middle'' decompactified Reissner-Nordstr{\"o}m space-time region and subsequently 
connecting the latter through the second ``throat'' to another ``right'' compactified 
Bertotti-Robinson universe, is {\em traversable} w.r.t. the {\em proper time} of a 
traveling observer. This property is similar to the {\em proper time} 
traversability of other symmetric and asymmetric (``one-throat'') wormholes produced by 
\textsl{LL-brane} sitting on the wormhole ``throat'' 
\ct{our-WH,rot-WH,ER-bridge,varna-09,BR-WH}.

Indeed, let us consider test particle (``traveling observer'') dynamics
in the asymmetric wormhole background given by \rf{left-BR}--\rf{right-BR}, which is
described by the action:
\be
S_{\mathrm{particle}} = \h \int d\l \Bigl\lb\frac{1}{e}\xdot^\m \xdot^\n G_{\m\n}
- e m_0^2 \rb \; .
% S_{\mathrm{particle}} = \int d\l \Bigl\lb\h\Bigl(\frac{1}{e}\xdot^\m \xdot^\n G_{\m\n}
% - e m_0^2\Bigr) - q_0 \xdot^\m \cA_\m\Bigr\rb \; .
\lab{test-particle}
\ee
Using energy $\cE$ and orbital momentum $\cJ$ conservation and introducing the 
{\em proper} world-line time $s$ ($\frac{ds}{d\l}= e m_0$), the ``mass-shell'' equation
(the equation w.r.t. the ``einbein'' $e$ produced by the action \rf{test-particle})
yields:
% \be
% \frac{1}{m_0^2} \(\cE - q_0 \cA_v (\eta)\)^2 = \(\frac{d\eta}{ds}\)^2 +
% A(\eta) \Bigl( 1 + \frac{\cJ^2}{m_0^2 C(\eta)}\Bigr) \; ,
% \lab{particle-eq-1}
% \ee
% where:
% \be
% \cA_v (\eta) = \frac{Q^2}{4\pi\,(r_0 + \eta)} \quad \mathrm{for} \;\; \eta >0
% \quad , \quad
% \cA_v (\eta) = \frac{\eta}{2\sqrt{\pi}r_0} + \frac{Q^2}{4\pi\,r_0} 
% \quad \mathrm{for} \;\; \eta <0 \; ,
% \lab{Maxwell-pot}
% \ee
% is the Maxwell potential ($\cF_{v\eta}= -\pa_\eta \cA_v$ with $\cF_{v\eta}$
% as in \rf{left-BR}--\rf{right-RN-1}),
% \be
% \(\frac{d\eta}{ds}\)^2 + A(\eta) \Bigl( 1 + \frac{\cJ^2}{m_0^2 C(\eta)}\Bigr) 
% = \frac{\cE^2}{m_0^2} \; ,
% \lab{particle-eq-1}
% \ee
% and the metric coefficients $\cA (\eta),\, C(\eta)$ are the same as in 
% \rf{left-BR}--\rf{right-RN-2}.
% Rewriting \rf{particle-eq-1} in the standard form:
\be
% \eta^{\pr\, 2} + \cV_{\mathrm{eff}} (\eta) = \frac{\cE^2}{m_0^2}
\(\frac{d\eta}{ds}\)^2 + \cV_{\mathrm{eff}} (\eta) = \frac{\cE^2}{m_0^2}
\quad ,\quad 
\cV_{\mathrm{eff}} (\eta) \equiv A(\eta) \Bigl( 1 + \frac{\cJ^2}{m_0^2 C(\eta)}\Bigr) 
\lab{particle-eq-2}
\ee
where the metric coefficients $\cA (\eta),\, C(\eta)$ are given in
\rf{left-BR}--\rf{right-BR}.
% with ``effective potential'':
% \br
% \cV_{\mathrm{eff}} (\eta) \equiv 
% \frac{\eta^2}{r_0^2}\Bigl\lb 1 + \frac{\cJ^2}{m_0^2 r_0^2} -
% \frac{q_0^2}{4\pi\,m_0^2}\Bigr\rb + 
% \frac{\eta}{r_0}\,\frac{q_0}{m_0^2 \sqrt{\pi}}\(\cE - \frac{q_0 Q^2}{4\pi\,r_0}\)
% + \frac{q_0 Q^2}{2\pi\, r_0 m_0^2} \(\cE - \frac{q_0 Q^2}{8\pi\,r_0}\)
% \quad \mathrm{for} \; \eta <0 \; ,
% \lab{eff-pot-BR} \\
% \cV_{\mathrm{eff}} (\eta) \equiv
% A_{\mathrm{RN}} (r_0 + \eta) \Bigl\lb 1 + \frac{\cJ^2}{m_0^2 (r_0 + \eta)^2}\Bigr\rb
% + \frac{q_0 Q^2}{2\pi\,m_0^2 (r_0 + \eta)} \Bigl\lb \cE  
% -  \frac{q_0 Q^2}{8\pi (r_0 + \eta)}\Bigr\rb \quad
% \mathrm{for} \; \eta >0 \; .
% \lab{eff-pot-RN}
% \er
% Here $A_{\mathrm{RN}}$ is the same as in \rf{RN-standard} and $r_0, m, Q^2$ are
% determined in \rf{param-1}, \rf{param-2}--\rf{param-3}.

\begin{figure}
\begin{center}
\includegraphics[width=12cm,keepaspectratio=true]{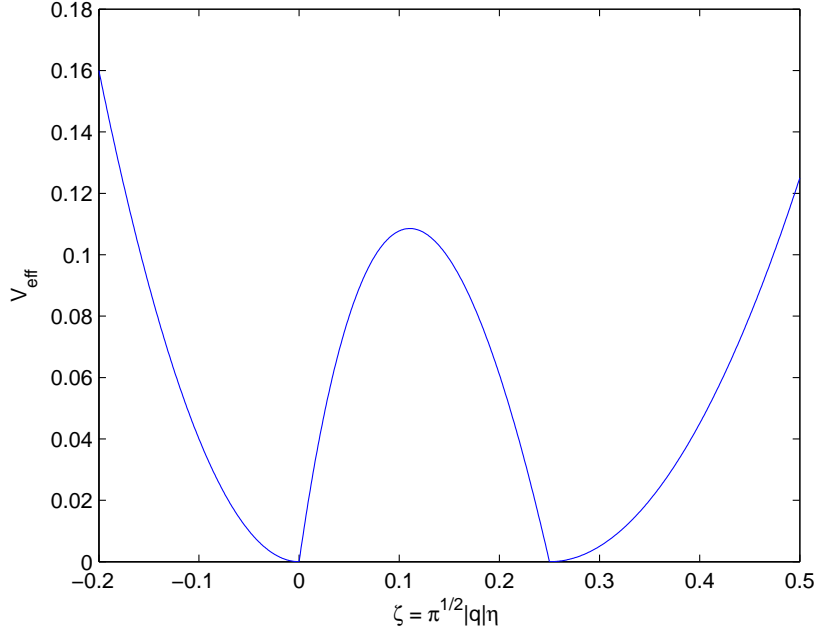}
\caption{Shape of $\cV_{\mathrm{eff}}(\eta)$ as a function of the rescaled coordinate 
$\zeta = \sqrt{\pi}|{\bar q}|\eta$}
\end{center}
\end{figure}

The shape of the ``effective potential'' is depicted on Fig.1 in the simple
case of $\cJ=0$, \textsl{i.e.}, for a purely ``radial'' motion along the
$\eta$-direction. In particular, in both ``left'' and ``right'' Bertotti-Robinson
universes it has harmonic-oscillator-type form.
% For generic values of the parameters the effective potential in the Bertotti-Robinson
% universe \rf{left-BR} (\textsl{i.e.}, for $\eta <0$) 
% has harmonic-oscillator-type form. 
% Therefore, a ``radially'' moving traveling observer % ($q_0=0$),
% starting in the Reissner-Nordstr{\"o}m universe \rf{right-RN-1} (\textsl{i.e.},
% at some $\eta >0$) and moving ``radially'' along the $\eta$-direction towards the
% horizon, will cross the wormhole ``throat''
% ($\eta=0$) within finite interval of his/her proper time, then will continue into the 
% Bertotti-Robinson universe subject to harmonic-oscillator deceleration force, will
% reverse back at the turning point and finally will cross the ``throat'' back into the
% Reissner-Nordstr{\"o}m universe.
Therefore, a ``radially'' moving traveling observer (with sufficiently large
energy $\cE$) will periodically ``shuttle'' along the $\eta$-direction within 
{\em finite proper-time} intervals between the two turning points $\eta_{(-)} <0$ 
in the ``left'' Bertotti-Robinson universe and $\eta_{(+)} > {\bar r}_0 - r_0$ 
in the ``right'' Bertotti-Robinson universe, where 
$\cV_{\mathrm{eff}} (\eta_{(\pm)}) = \cE^2 / m_0^2$.

% %\vspace{.2in}

% % %%%%%%%%%%%%%%%%%%%%%%%%%%%%%%%%%%%%%%%%%%%%%%%%%%%%%%%%%%%%%%%%%%
% % \%\%\%\%\%\%\%\%\%\%\%\%\%\%\%\%\%\%\%\%\%\%\%\%\%\%\%\%\%\%\%\%\%

% % \textbf{More COMMENTS - ``acceleration'' etc. for CHARGED test particles}

% % Effective force around the ``throat'':
% % \br
% % \partder{\cV}{\eta}\bv_{+0} = 16\pi |\chi| \Bigl( 1+\frac{\cJ^2}{m_0^2 r_0^2}\Bigr)
% % - \frac{q_0 Q^2}{2\pi\, r_0^2 m_0^2} \(\cE - \frac{q_0 Q^2}{4\pi\,r_0}\)
% % \lab{eff-force-1} \\
% % \partder{\cV}{\eta}\bv_{-0} =
% % \frac{q_0}{m_0^2 r_0\sqrt{\pi}} \(\cE - \frac{q_0 Q^2}{4\pi\,r_0}\)
% % \lab{eff-force-2}
% % \er
% % Note $\cV (0)=\frac{q_0 Q^2}{2\pi\, r_0 m_0^2} \(\cE -\frac{q_0 Q^2}{8\pi\,r_0}\)$, 
% % therefore:
% % \be
% % \(\frac{d\eta}{ds}\)^2 + \cV^{(1)} (\eta) = 
% % \frac{1}{m_0^2} \Bigl(\cE - \frac{q_0 Q^2}{4\pi\,r_0}\Bigr)^2 \quad ,\quad
% % \cV^{(1)} (\eta) \equiv \cV (\eta) - \cV (0)
% % % = \eta\,\partder{\cV}{\eta}\bv_{\pm 0} + O(\eta^2)
% % \lab{particle-eq-3}
% % \ee

% % \%\%\%\%\%\%\%\%\%\%\%\%\%\%\%\%\%\%\%\%\%\%\%\%\%\%\%\%\%\%\%\%\%
% % %%%%%%%%%%%%%%%%%%%%%%%%%%%%%%%%%%%%%%%%%%%%%%%%%%%%%%%%%%%%%%%%%%

% % \vspace{.2in}

On the other hand let us point out that, as in the case of the previously constructed 
symmetric and asymmetric wormholes via \textsl{LL-branes} sitting on their ``throats'' 
\ct{our-WH,rot-WH,ER-bridge,varna-09,BR-WH}, the present ``two-throat'' wormhole is
{\em not} traversable w.r.t. the ``laboratory'' time of a static observer in
either of the three universes.

% %%%%%%%%%%%%%%%%%%%%%%%%%%%%%%%%%%%%%%%%%%%%%%%%%%%%%%%%%%%%%
% %% ROLE OF LL-BRANES TO CONSTRUCT NON-SINGULAR BLACK HOLES \ct{reg-BH}
Let us also mention the crucial role of \textsl{LL-branes} in
constructing non-trivial examples of {\em non-singular} black holes,
\textsl{i.e.}, solutions of Einstein equations with black hole type geometry
in the bulk space-time, in particular possessing horizons, but with {\em no}
space-time singularities in the center of the geometry. For further details
we refer to \ct{reg-BH}, where a solution of the Einstein-Maxwell-Kalb-Ramond
system coupled to a charged \textsl{LL-brane} has been obtained describing a 
{\rm regular} black hole. The space-time manifold of the latter consists of 
de Sitter interior region and exterior Reissner-Nordstr{\"o}m region glued
together along their common horizon (it is the {\em inner} horizon from the 
Reissner-Nordstr{\"o}m side).

% %%%%%%%%%%%%%%%%%%%%%%%%%%%%%%%%%%%%%%%%%%%%%%%%%%%%%%%%%%%%%
Finally, let us point out one possible direction for generalizing the present
results to higher-dimensional cases. Notice that one generalization of
Bertotti-Robinson solutions in $D\!=\!6$ is the famous Freund-Rubin solution
\ct{freund-rubin} of the form $AdS_4 \times S^2$. Another generalization of the form
$Minkowski_4 \times S^2$ was found in \ct{Dav-Ed-89} upon introducing a
bare cosmological constant in $D\!=\!6$. One could envisage a construction of a
$D\!=\!6$-dimensional wormhole-like solution of gravity/gauge-field system
self-consistently interacting with two (widely separated) codimension-one
lightlike branes describing a transition from a ``left'' Freund-Rubin-type
universe via ``intermediate'' $D\!=\!6$ Reissner-Nordstr{\"o}m-de-Sitter region
(cf. \ct{myers-perry}) to a different ``right'' Freund-Rubin-type universe
where the lightlike 4-branes occupy the interfaces between the various
universes.

%%%%%%%%%%%%%%%%%%%%%%%%%%%%%%%%%%%%%%%%%%%%%%%%%%%%%%%%%%%%
%%%%%%%%%%%%%%%%%%%%%%%%%%%%%%%%%%%%%%%%%%%%%%%%%%%%%%%%%%%%%
\begin{acknowledgements}
E.N. and S.P. are supported by Bulgarian NSF grant \textsl{DO 02-257}.
E.G. thanks the astrophysics and cosmology group at PUCV, Chile,  for hospitality.
Also, all of us acknowledge support of our collaboration through the exchange
agreement between the Ben-Gurion University of the Negev and the Bulgarian Academy 
of Sciences.
\end{acknowledgements}

%%%%%%%%%%%%%%%%%%%%%%%%%%%%%%%%%%%%%%%%%%%%%%%%%%%%%%%%%%%%%

\end{document}